\documentclass[11pt,a4paper]{scrartcl}
\pdfoutput=1
\usepackage{CLICdp}

\usepackage{CLICdp_definitions}
\usepackage{float}

\renewcommand{\Pep}{\ensuremath{e^{+}}\xspace}
\renewcommand{\Pem}{\ensuremath{e^{-}}\xspace}

\renewcommand{\Pem}{$\mathrm{e^{-}}$\xspace}
\renewcommand{\Pep}{$\mathrm{e^{+}}$\xspace}
\renewcommand{\PGg}{$\mathrm{\gamma}$\xspace}
\renewcommand{\PGn}{$\mathrm{\nu}$\xspace}
\renewcommand{\PAGn}{$\mathrm{\bar{\nu}}$\xspace}
\newcommand{\PPem}{\ensuremath{ P_{e^{-}} }\xspace}
\newcommand{\PPep}{\ensuremath{ P_{e^{+}} }\xspace}

\newcommand{\RA}{\rightarrow}
\usepackage{grffile}  

\title{Physics performance for Dark Matter searches at $\sqrt{s}=$ 3 TeV at
CLIC using mono-photons and polarised beams.}

\clicdpnote{2021}{001}  

\date{\today}

\addauthor{J-J.~Blaising\thanks{jean-jacques.blaising@cern.ch}}{\institute{1}}
\addauthor{P.~Roloff}{\institute{2}}
\addauthor{A.~Sailer}{\institute{2}}
\addauthor{U.~Schnoor}{\institute{2}}

\addinstitute{1}{Laboratoire d'Annecy-le-Vieux de Physique des Particules, Annecy-le-Vieux, France}
\addinstitute{2}{CERN, Switzerland}

\onbehalfof{the CLICdp Collaboration} 

\abstract{
At \Pem \Pep colliders, Weakly Interacting Massive Particles (WIMPs) are candidates for dark matter (DM)
and can be searched for using as tag a photon from initial state radiation. 
The potential for detecting DM at the Compact Linear Collider (CLIC) is investigated
at \mbox{$\sqrt{s}=$ 3 TeV}.
The sensitivity of the search is estimated by computing the 95\% confidence level
upper limit cross section as a function of the dark matter mass.
Left-handed (right-handed) polarised \Pem beams increase (decrease) respectively 
the Standard Model backgrounds  
and are essential to characterize the WIMPs properties and control the systematic errors. 
Using right-handed polarised \Pem beams is decreasing significantly the 95\% confidence level
cross section. 
Using the ratio of the energy distributions for left-handed and right-handed polarised \Pem beams, 
systematic errors cancel out. Computing the 95\% confidence level upper limit cross section
using the ratio requires a model assumption to compute the expected number of signal events.
Exclusion limits for dark matter are derived using dark matter Simplified Models 
for two values of the e-e-mediator vertex coupling, a mediator width of 10 GeV and
for a fixed value of the mediator-DM-DM coupling.  
For a mediator mass of 3.5 TeV,
the measurement of the differential distribution of the significance as a function of
the photon energy
for the process \mbox{\Pem \Pep $\to$ X X \PGg}
allows the discrimination between different dark matter mediators and  
the measurement of the WIMP mass to nearly half the centre-of-mass energy.
For a \mbox{1 TeV} WIMP, the mass is determined with a 1\% accuracy. 
}

\titlecomment{This work was carried out in the framework of the CLICdp Collaboration} 

\graphicspath{ {./logos/}{./figures/} {./plots/} }

\begin{document}

\titlepage

\newcommand{\latex}{\LaTeX\xspace}
\lstset{defaultdialect=[LaTeX]TeX}
%
\message{**:Start definitions}
\def\Bi {\tilde{B}}
\def\Wi {\tilde{W}}
\def\Hi {\tilde{H}}
\def\susy#1{\ensuremath{\tilde{\mathrm{#1}}}}%
\def\slepton   #1{\ensuremath{\susy{\ell}^{#1}}}
\def\selectron #1{\ensuremath{\susy{e}^{#1}}}
\def\smuon     #1{\ensuremath{\susy{\mu}^{#1}}}
\def\stau      #1{\ensuremath{\susy{\tau}^{#1}}}
\def\sneutrino {\susy{\nu}}
\def\chargino  #1{\ensuremath{\susy{\chi}_1^{#1}}}
\def\charginotwo  #1{\ensuremath{\susy{\chi}_2^{#1}}}
\def\charginos#1{\ensuremath{\susy{\chi}_{1,2}^{#1}}}
\def\neutralino#1{\ensuremath{\susy{\chi}_{#1}^0}}
\def\charginop#1{\ensuremath{\susy{\chi}_{#1}^+}}
\def\charginom#1{\ensuremath{\susy{\chi}_{#1}^-}}
\section{Introduction}
The Compact Linear Collider (CLIC) is a mature option for a future TeV-scale \Pem \Pep collider.
CLIC is proposed to run according to a staging scenario with different centre-of-mass energies,
\mbox{$\sqrt{s}$=380 GeV}, 1.5 TeV and 3 TeV. The baseline machine allows for up to 
$\pm$ 80\% electron polarisation.
The lowest energy stage will concentrate on Higgs and Top physics
~\cite{HiggsPhys:}, ~\cite{TopPhys:} collecting data corresponding to an integrated luminosity 
of \SI{1}{ab^{-1}}~\cite{Robson1}.

At all centre-of-mass energies, Beyond Standard Model (BSM) searches will also be performed,
among which are searches for dark matter (DM).
The existence of DM has been firmly established through observation of its gravitational effects,
but its nature is not established.  
A widely accepted hypothesis on the form for dark matter is that it is composed of weakly 
interacting massive particles (WIMPs) that interact only through gravity and the weak force.
There are alternative hypotheses that attempt to account for the observations without 
invoking additional matter.
WIMP searches at \Pem \Pep colliders are complementary to the direct
and indirect searches and to the hadron collider searches. 
As the WIMP escapes detection, in \Pem \Pep collisions the signature is a photon 
radiated off the initial-state leptons and missing energy. 
Searches for single and multiphoton final states with missing energy have 
been performed by LEP experiments 
~\cite{SinglePhoton:A}, ~\cite{SinglePhoton:B}
~\cite{SinglePhoton:C},
~\cite{SinglePhoton:D}
and investigated at the International Linear Collider 
~\cite{ILC:A}.
The main background processes are 
\mbox{\Pem \Pep $\to$ \PGn \PAGn \PGg} and
\mbox{\Pem \Pep $\to$ \Pem \Pep \PGg}.
The neutrino background is irreducible, but can be enhanced or reduced by
changing the \Pem beam polarisation. 
Radiative Bhabha scattering has a large cross section and mimics the
signal if both leptons are undetected. For the reduction of this background, excellent
hermiticity in the forward region of the detector is required. 
%
To assess the CLIC physics potential, the 95\% confidence level upper limit 
on the cross section is determined as a function of dark matter mass
for different polarisation conditions.
The 95\% confidence level upper limit cross section as a function of dark matter mass
is then used to derive exclusion limits 
in the framework of dark matter Simplified Models 
(DMSMs) 
\cite{DMSMs:}, ~\cite{DMSMsA:},  ~\cite{Wulzer:}.
Simplified models are designed to grab the basic details
of collider phenomenology found in rigorously-derived
theories of new physics, without the complexity of the full theory. 
They are used to compare hadron and lepton collider physics potential.
These simplified models
assume that DM is a Dirac fermion X and there is an additional
heavy particle mediating the SM-DM interaction, the \textquotedblleft mediator Y \textquotedblright.
In the most basic set of these models, the mediator is a vector,
an axial-vector, a scalar or a pseudo-scalar boson. 
In this study, CLICdp focused on the subset of the models where the
mediator is exchanged in the s-channel. 
In the presence of the signal, using the photon energy distribution, 
the distribution of the significance $Z$ as a function of $E_{\gamma}$, $dZ/dE_{\gamma}$ is computed
for pseudo-data and for different mediator templates coupling to dark matter. 
The $\chi^{2}$ between the significance of pseudo-data and templates is computed.
It is used to discriminate among different mediator hypotheses and to determine the dark matter mass.
\section{Dark Matter exclusion limits at 3 TeV}
\subsection{Event simulation and selection  at 3 TeV}
Cross-section calculation and event generation are done using 
the WHIZARD 2 program
~\cite{Whizard:2008}.
The cross sections are calculated at 3 TeV without and with \Pem beam polarisation.
Beamstrahlung effects on the luminosity spectrum are included using results of the CLIC beam simulation
for the 3 TeV accelerator parameters.
There are three sources of the centre-of-mass energy spread: the momentum spread in the linac,
the beamstrahlung which creates a long tail, and initial state radiation (ISR).
The first two are collectively refered to as ``luminosity spectrum''.
The luminosity spectrum is obtained from the
GuineaPig~\cite{Guineapig:} beam simulation, it is interfaced to WHIZARD using 
circe2~\cite{Whizard:2008}. 
The standard procedure to take ISR effects into account when generating events with WHIZARD 
is to use the built-in lepton ISR structure function which includes all orders of soft and 
soft-collinear photons as well as up to the third order in high-energy collinear photons. 
However, this approach allows only for a proper modelling of the kinematics of the
hard scattering, but is not suitable when we expect photons to be detected in the experiment.
For proper description of the photon kinematics, hard non-collinear photon emission
should be included in the generation of the background processes on the matrix element level.
To obtain a realistic distribution of the photon polar angle, up to three photons are included in 
the matrix element.
Events with ISR photons emitted in the same kinematic region as matrix element (ME) photons are 
rejected using a merging procedure ~\cite{PAWEL:A}. 
Table~\ref{tab:3000CrossSections} shows the integrated luminosity $(L)$
assumptions at \mbox{$\sqrt{s}$=3 TeV} 
for different beam polarisation conditions.
It shows also the cross sections of the two main
Standard Model processes for different beam polarisation conditions requiring
$Pt_{\gamma}/\sqrt{s} >$ 0.02, \mbox{$ 10^\circ < \theta_{\gamma} < 170^\circ $} and
without ISR/ME merging cut.
$\theta_{\gamma}$ and $Pt_{\gamma}$ are the polar angle and the transverse momentum of the 
photon respectively.
\PPem is the longitudinal degree of polarisation of the \Pem beam, 
\PPem > 0 (right-handed polarisation)
No additional selection cut is applied for the cross section calculation and event generation. 
The ISR/ME merging efficiency is 70\% for \mbox{\PGn \PAGn \PGg}
and 69\% for \mbox{\Pem \Pep \PGg}.
\begin{table} [!htbp]
\centering
\caption{ Cross sections of main Standard Model background processes at $\sqrt{s}$ = 3 TeV
}
\begin{tabular}{ l c c c }
\hline
 Polarisation                      & No                  & \PPem,\PPep  & \PPem,\PPep \\
                                   &                     & -80~,~0~~~             & +80~,~0~~~~~     \\
 Integrated luminosity $(L)$       & 5 $\SI{}{ab^{-1}}$  & 4 $\SI{}{ab^{-1}}$ & 1 $\SI{}{ab^{-1}}$ \\
$\sigma$(\Pem \Pep $\to$  \PGn \PAGn \PGg (\PGg)) [fb] & $1.06\times10^{3}$  & $1.88\times10^{3}$ & $2.35\times10^{2}$ \\
$\sigma$(\Pem \Pep $\to$ \Pem \Pep \PGg (\PGg)) [fb] & $1.92\times10^{3}$  & $1.96\times10^{3}$ & $1.89\times10^{3}$ \\ 
\end{tabular}
\label{tab:3000CrossSections}
\end{table}

A fast simulation is used to
compute the expected \mbox{\Pem \Pep $\to$ \PGn \PAGn \PGg (\PGg)}
and \mbox{\Pem \Pep $\to$ \Pem \Pep \PGg (\PGg)} backgrounds.
To take into account the detector resolution, the momentum and energy of the particles are
randomly modified using Gaussian resolution parameters according to the particle type.
Table~\ref{tab:3000Parameters} shows the angular coverage of the different detectors
used for the analysis, the simulation parameters 
and parameter values used for the fast simulation of electrons and photons.
$\theta$ is the polar angle of the particle, electron or photon.
The parameter values are obtained from full simulation studies.
Details about CLIC event simulation and detector performance can be found in ~\cite{cdrvol2},
~\cite{Sailer} and methods of the background simulation and electron 
reconstruction are described in ~\cite{AndrePhD:} and
~\cite{Sailer1}.
%
\begin{table} [!htbp]
\centering
\caption{Detector regions, simulation parameters and efficiencies at $\sqrt{s}$ = 3 TeV }
\begin{tabular}{ c c c c c c}
\hline
  Detector          & Angular region      & Particle     & $Emin$ & $\sigma(E)/E $  & $\epsilon_{D}$ \\
                    & mrad                &              & GeV    & $E~\SI{}{[GeV]}$           &  \%    \\
\hline
  BeamCal           & $15 < \theta < 40$  & e, \PGg & 1000   & 0.1            & $f(E,\theta)$ \\
  LumiCal           & $40 < \theta < 100$ & e, \PGg &  500   & $0.8/\sqrt{E}$ & 99      \\
  ECal No tracking  & $100 < \theta < 175$& e, \PGg &   50   & $0.2/\sqrt{E}$ & 99      \\
  ECal and tracking & $\theta > 175$      & e       &   20   & F(~\ref{equ:resolT}) & 99.5    \\
  ECal and tracking & $\theta > 175$      & \PGg      &   20   & $0.17/\sqrt{E}$          & 99    \\
\end{tabular}
\label{tab:3000Parameters}
\end{table}

$Emin$ is the minimum energy required to compute the energy of the particle and apply
a detection efficiency. 
$\sigma(E)/E $ is the energy resolution, and $\epsilon_{D}$ is the detection efficiency.
In the BeamCal, \mbox{$\sigma(E)/E $ =0.1~[\SI{}{GeV}]} 
and the detection efficiency is a function of the energy and of the angle of the particle.
The efficiency $f(E,\theta)$ is computed using the ElectronEfficiency library developed for
the CLIC CDR~\cite{ElecEffiLib:}.  
In the LumiCal, $\sigma(E)/E $ = $0.8/\sqrt{E}$ ~[\SI{}{GeV}]  and the detection efficiency is $\epsilon_{D}$=99\%.
In the Ecal region without tracking system, $\sigma(E)/E $ = $0.2/\sqrt{E}$ ~[\SI{}{GeV}]
and the detection efficiency is $\epsilon_{D}$=99\%.
In the "signal region", $10^\circ < \theta < 170^\circ $, 
the energy resolution and the detection efficiency are
different for electrons and photons.
For the electrons, the momentum resolution function is:
\begin{eqnarray}
\label {equ:resolT}
\frac{\sigma(P)}{P} \approx a \cdot P \oplus b \cdot \frac{1}{\sqrt{\sin\theta}} \oplus c \cdot \frac{\cos\theta}{\sin\theta},
\end{eqnarray}
the parameter a, b and c represents the contribution from the curvature measurement,
from the multiple-scattering and from the angular resolution respectively.
The values were obtained by fitting the $P_{T}$ resolution of fully simulated charged
particle data as a function of the $P_{T}$~\cite{cdrvol2}.
For $\theta=90^{\circ}$, 
$a=2.0\times 10^{-5}~[\SI{}{GeV^{-1}}]$, 
$b=2.0\times 10^{-3}$ and $c=2.0\times 10^{-4}$.
For tracks measured in the tracker barrel $\theta>40^{\circ}$ and $\theta<140^{\circ}$
parameter $a$ is independent of $\theta$.
For tracks measured in the end cap disks the path length L depends on $\theta$ and
therefore $a$ depends on $\theta$.
The detection efficiency is $\epsilon_{D}$=99.5\%. 
For the photons, $\sigma(E)/E $ = $0.17/\sqrt{E}$ ~[\SI{}{GeV}] and the detection efficiency is $\epsilon_{D}$=99\%. 
With these parameters, listed in Table~\ref{tab:3000Parameters}, 
the fake rate due to beam-induced backgrounds, is well below $10^{-4}$ in all detector regions.

\begin{figure} [!htbp]
  \centering
  \begin{subfigure}[b]{0.48\textwidth}
    \includegraphics[width=\textwidth]{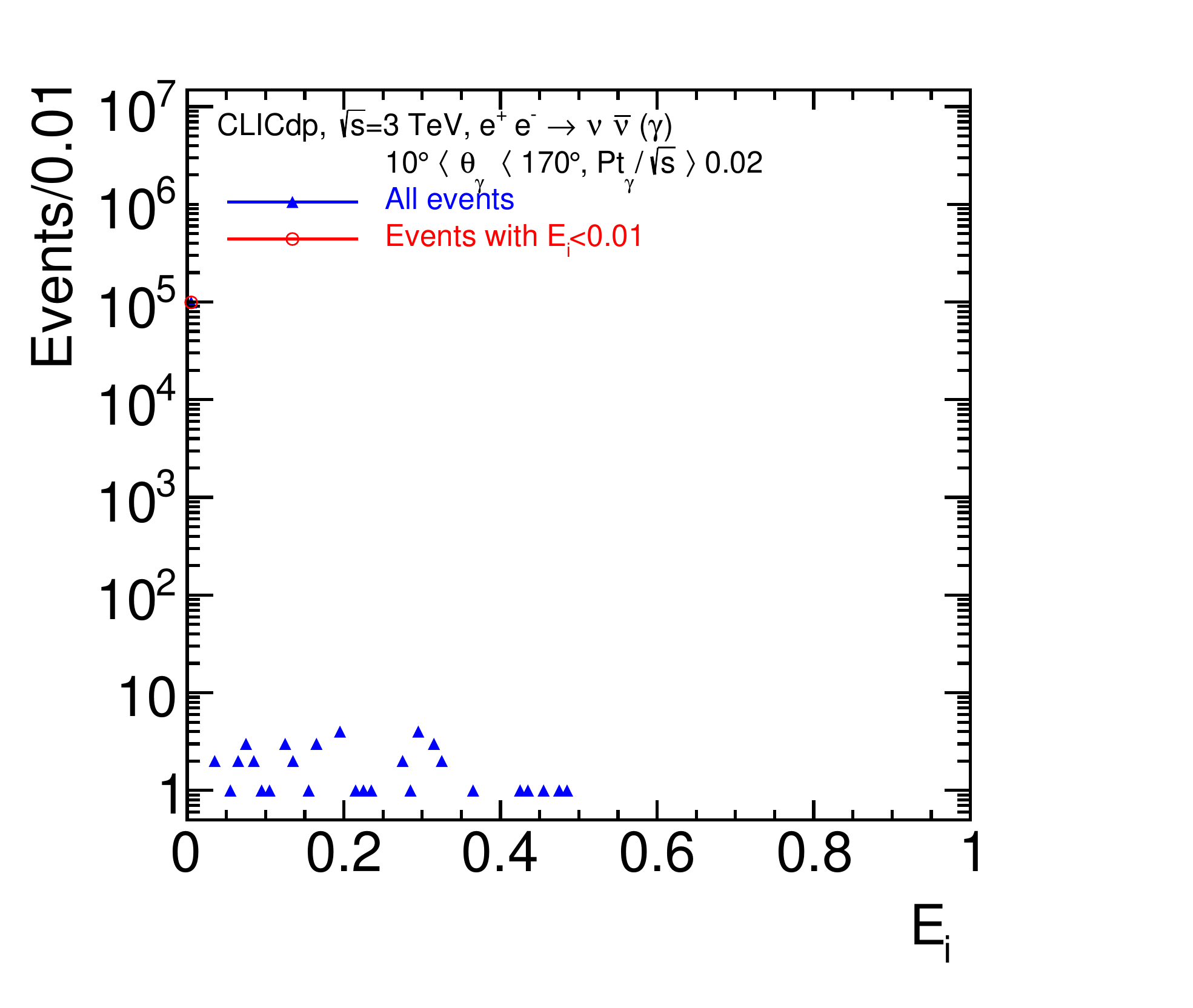}
    \caption{Photon isolation selection distribution, $dN/dE_{i}$ }
    \label{fig:h1EinN}
  \end{subfigure}
  \hfill
  \begin{subfigure}[b]{0.48\textwidth}
    \includegraphics[width=\textwidth]{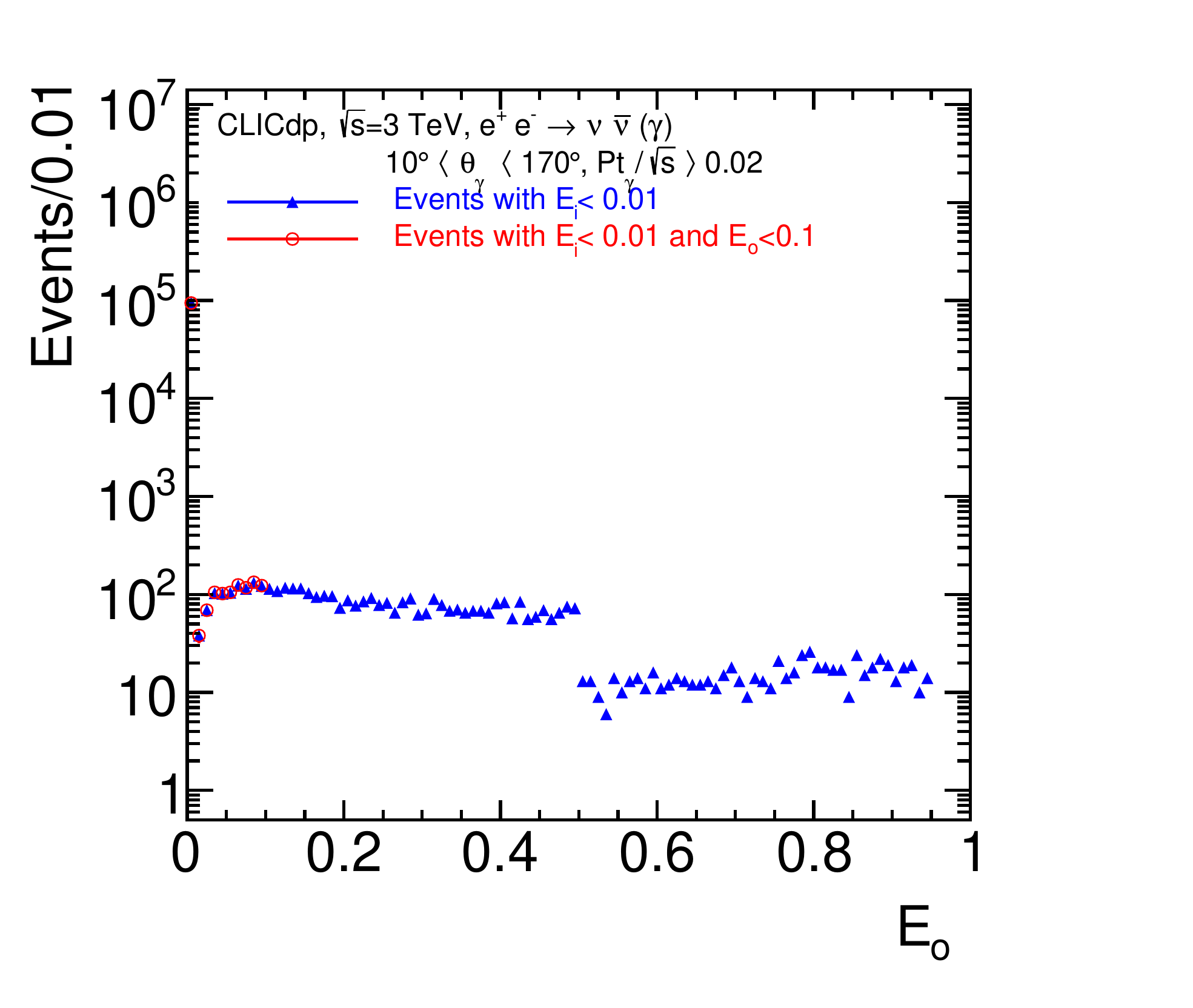}
    \caption{Missing energy selection distribution $dN/dE_{o}$}
    \label{fig:h1EonN}
  \end{subfigure}
  \caption{ Mono-photon selection 
    for the process \mbox{ \Pem \Pep $\to$ \PGn \PAGn \PGg (\PGg)}
   \subref{fig:h1EinN} Photon isolation selection distribution $dN/dE_{i}$    
   \subref{fig:h1EonN} Photon missing energy selection distribution $dN/dE_{o}$ }
\end{figure}

\begin{figure} [!htbp]
  \centering
  \begin{subfigure}[b]{0.48\textwidth}
    \includegraphics[width=\textwidth]{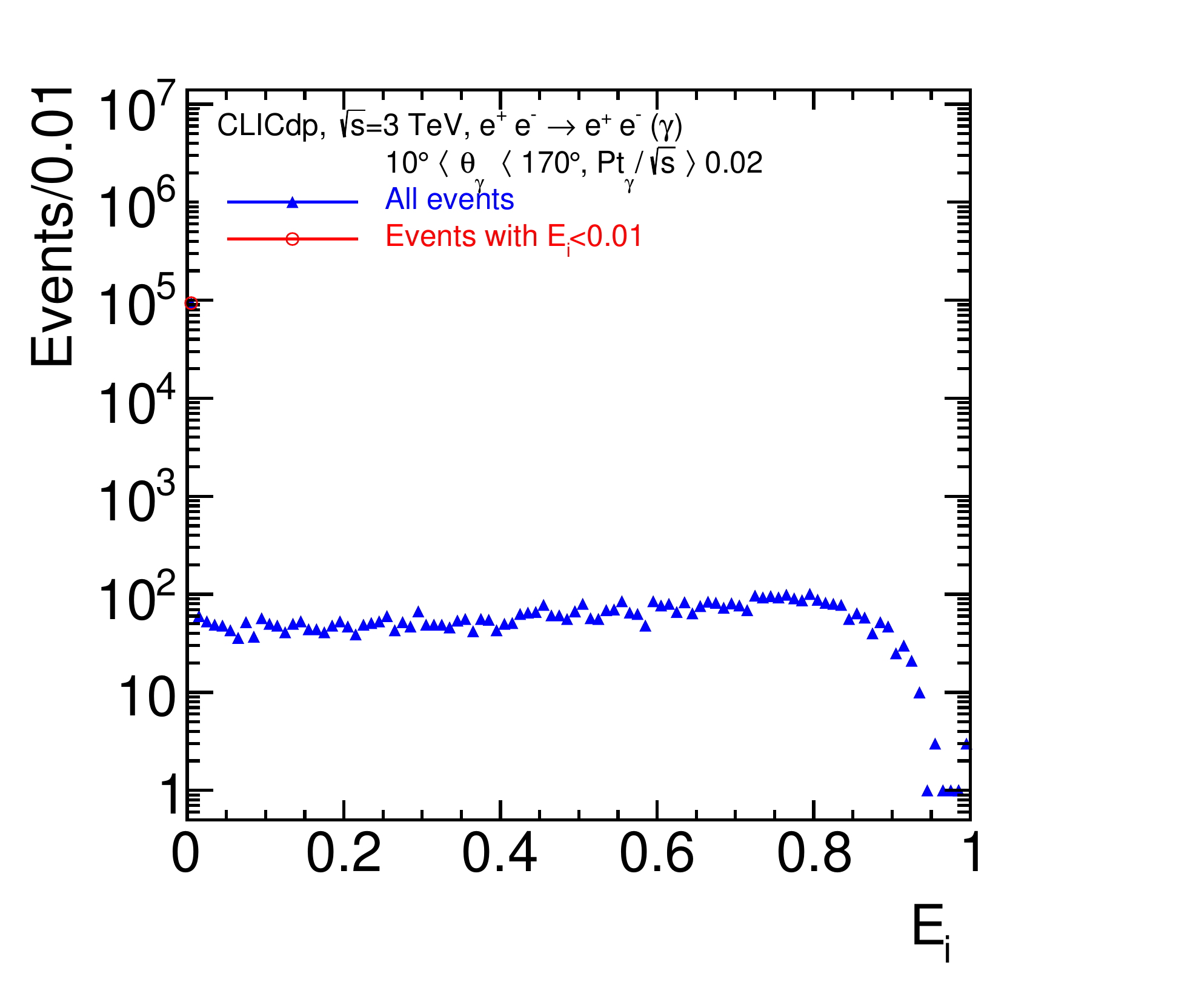}
    \caption{Photon isolation selection distribution, $dN/dE_{i}$ }
    \label{fig:h1EieE}
  \end{subfigure}
  \hfill
  \begin{subfigure}[b]{0.48\textwidth}
    \includegraphics[width=\textwidth]{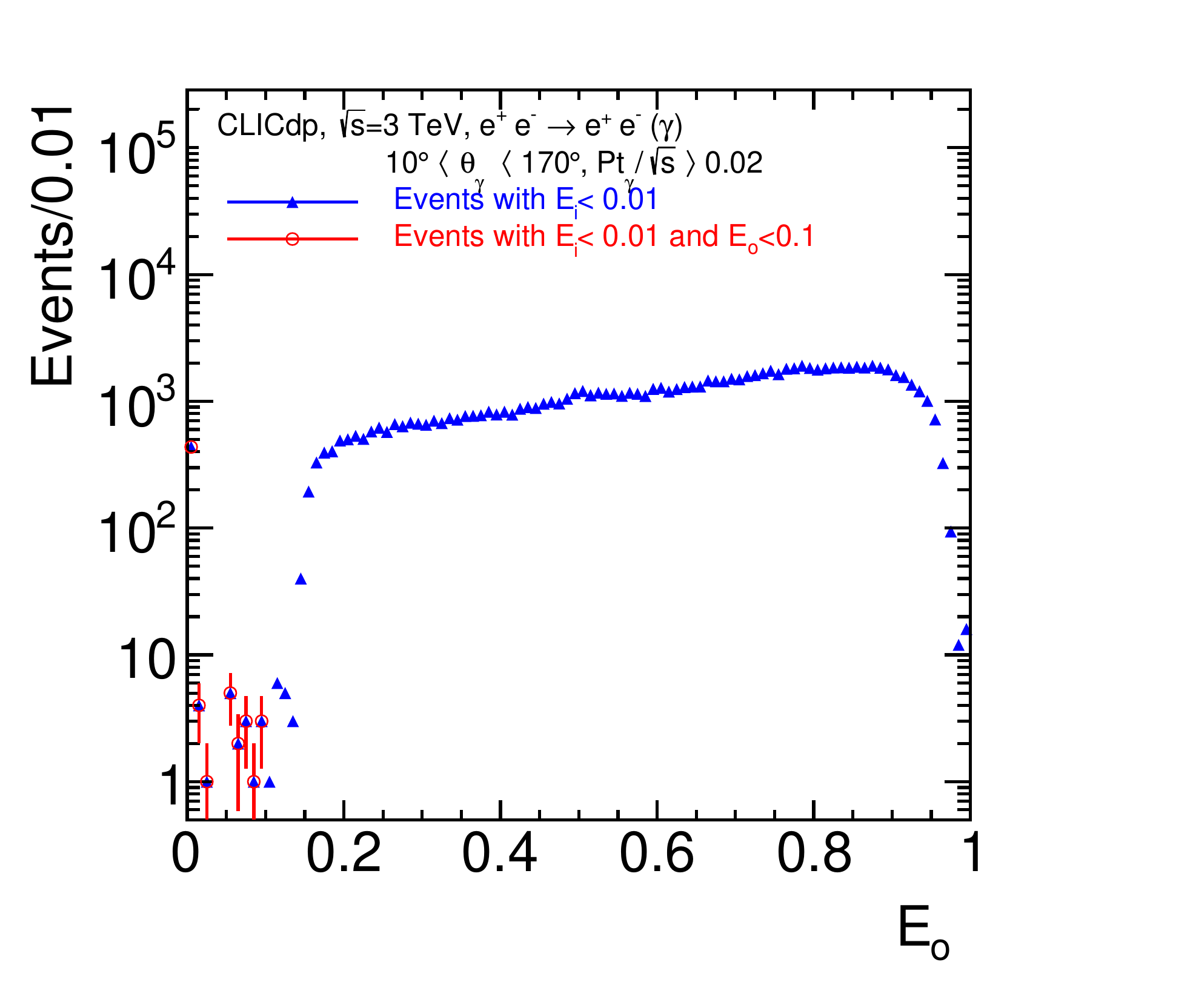}
    \caption{Missing energy selection distribution $dN/dE_{o}$}
    \label{fig:h1EoeE}
  \end{subfigure}
  \caption{ Mono-photon selection for the process
    \mbox{ \Pem \Pep $\to$ \Pem \Pep \PGg (\PGg)}
   \subref{fig:h1EieE} Photon isolation selection distribution $dN/dE_{i}$
   \subref{fig:h1EoeE} Photon missing energy selection distribution $dN/dE_{o}$ }
\end{figure}

The mono-photon selection requires an isolated photon with an energy greater than
60 GeV in the signal region and little energy in the
other parts of the detector. This selection includes:
\begin {itemize}
\item
A photon isolation selection based on the observable $E_{i}$ defined as:
\begin{eqnarray}
E_{i}= | E_{Tr} - E_{\gamma} | / E_{Tr}.
\end{eqnarray}
$E_{\gamma}$ is the photon energy and 
$E_{Tr}$ is the energy sum of the particles in a cone of width $\Delta R$=0.4 
around the photon direction with
\begin{eqnarray}
\Delta R= \sqrt{\Delta \eta^{2} + \Delta \phi^{2}}.
\end{eqnarray}
\item
A missing energy selection requiring little energy in addition to the photon energy 
\begin{eqnarray}
E_{o}= | E_{Tot} - E_{\gamma} | / E_{Tot}.
\end{eqnarray}
$E_{Tot}$ is the total energy of the event.
\end {itemize}
When there are several photons in the signal region the photons are sorted by energy.
The energy of the most energetic photon is considered to be the energy $E_{\gamma}$ of the 
signal photon and $E_{\gamma}$ is randomly modified using the photon energy resolution of the 
signal region.
The particles outside the signal region cannot be identified and are considered as electrons.
To compute $E_{Tr}$ and $E_{Tot}$, the energy of the remaining particles are considered if 
their energy is greater than $Emin$.
For all these particles the energy is randomly modified using the energy resolution of the
detector region where the particle is detected and the detection efficiency is taken into
account when computing $E_{Tr}$ and $E_{Tot}$.  
%
Figure~\ref{fig:h1EinN} shows for the process
\mbox{ \Pem \Pep $\to$ \PGn \PAGn \PGg (\PGg)},
the distribution $dN/dE_{i}$ of the observable $E_{i}$ used for the photon isolation selection.
The events with $E_{i}$>0.01 are events for which
the energy of an additional photon is measured in the cone around  
the high energy signal photon. 
Figure~\ref{fig:h1EonN} shows the distribution
$dN/dE_{o}$ of the observable $E_{o}$ used for the missing energy selection.
The events with $E_{o}$>0.1 are events for which
the energy of an additional photon is measured, either in the signal region  
or outside the signal region.   
Figure~\ref{fig:h1EieE}
and Figure~\ref{fig:h1EoeE}
show the same distributions $dN/dE_{i}$ and $dN/dE_{o}$
for the process
\mbox{ \Pem \Pep $\to$ \Pem \Pep \PGg (\PGg)}.
%
%
%
The events with $E_{o}$>0.1 are events for which
the energy of an additional photon or electron is measured, either in the signal region  
or outside the signal region.   
Taking into account the high energy required for the signal photon, $E_{\gamma}>$ 60 GeV and the
low beam-induced fake rate in all detection regions, potential
energy deposits originating from beam-induced background are neglected when computing
the mono-photon selection efficiency. 
%
%
Table~\ref{tab:3000SelEffi} shows for the two main SM backgrounds
\mbox{\Pem \Pep $\to$ \PGn \PAGn \PGg (\PGg)} and
\mbox{\Pem \Pep $\to$ \Pem \Pep \PGg (\PGg)}, 
the selection efficiencies of the
photon isolation cut $E_{i}$ and of the missing energy selection cut $E_{o}$.
The selection efficiencies for left-handed 80\% polarised \Pem beam 
and for right-handed 80\% polarised \Pem beam are the same.
\begin{table} [!htbp]
\centering
\caption{Selection cuts and selection efficiencies of the two main SM backgrounds}
\begin{tabular}{ l c c c }
\hline
                    &                   & Process              & Process \\
                    &  & \Pem \Pep $\RA$ \PGn \PAGn \PGg (\PGg) & \Pem \Pep $RA$ \Pem \Pep \PGg (\PGg) \\
\hline
Cut name            & Cut value         & Selection efficiency & Selection efficiency \\
$E_{i}$             &  $E_{i}$<0.01                  & 0.99                   & 0.935                 \\
$E_{i}$ and $E_{o}$ &  $E_{i}$<0.01 and $E_{o}$<0.1  & 0.95                   & 4.6$\times 10^{-3}$                \\
\end{tabular}
\label{tab:3000SelEffi}
\end{table}

\begin{figure} [!htbp]
  \centering
  \begin{subfigure}[b]{0.48\textwidth}
    \includegraphics[width=\textwidth]{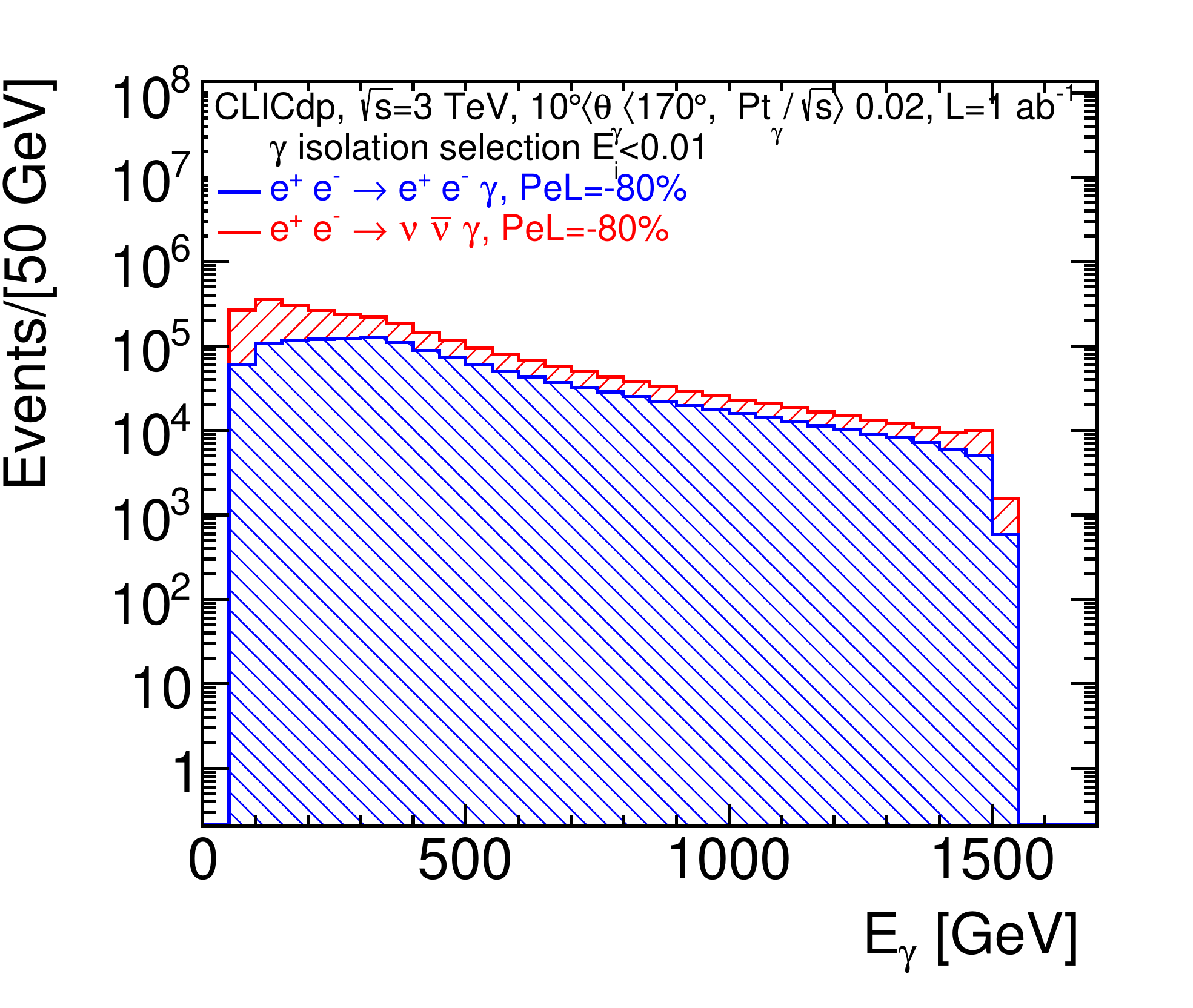}
    \caption{$dN/dE_{\gamma}$ spectrum with photon isolation selection $E_{i}<0.01$~~~~~~~~~~}
    \label{fig:1DStackB1N3000}
  \end{subfigure}
  \hfill
  \begin{subfigure}[b]{0.48\textwidth}
    \includegraphics[width=\textwidth]{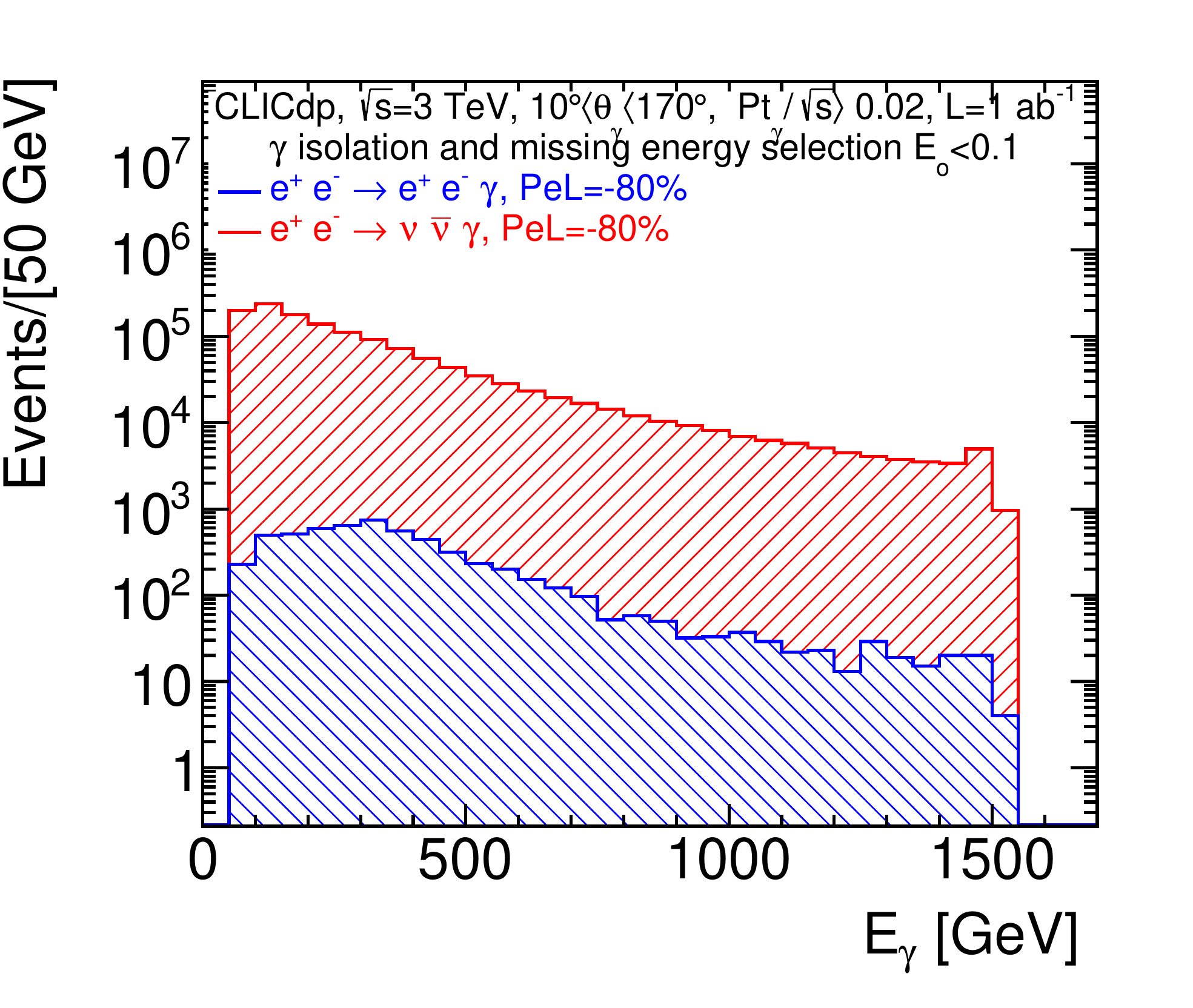}
    \caption{$dN/dE_{\gamma}$ spectrum with photon isolation selection $E_{i}<0.01$ and 
         missing energy selection $E_{o}<0.1$   }
    \label{fig:1DStackB13000}
  \end{subfigure}
  \caption{Left-handed 80\% polarised electron beam \mbox{ \Pem \Pep $\to$ \PGn \PAGn \PGg (\PGg)} and
   \mbox{ \Pem \Pep $\to$ \Pem \Pep \PGg (\PGg)} events with mono-photon selection 
   \subref{fig:1DStackB1N3000} with photon isolation selection $E_{i}<0.01$  
   \subref{fig:1DStackB13000} with photon isolation selections $E_{i}<0.01$ 
     and missing energy selection $E_{o}<0.1$  }
\end{figure}

Figure~\ref{fig:1DStackB1N3000} 
shows for a left-handed 80\% polarised \Pem beam (PeL), the stacked histogram of 
the photon energy distribution of 
\mbox{ \Pem \Pep $\to$ \PGn \PAGn \PGg (\PGg)}
and
\mbox{ \Pem \Pep $\to$ \Pem \Pep \PGg (\PGg)} 
events selected using the photon isolation selection $E_{i}<0.01$.
The number of events corresponds to an integrated luminosity of 1 $\SI{}{ab^{-1}}$.  
Figure~\ref{fig:1DStackB13000} shows the photon energy distribution
for events selected using
the photon isolation selection $E_{i}<0.01$ 
and the missing energy selection $E_{o}<0.1$.

\begin{figure} [!htbp]
  \centering
  \begin{subfigure}[b]{0.48\textwidth}
    \includegraphics[width=\textwidth]{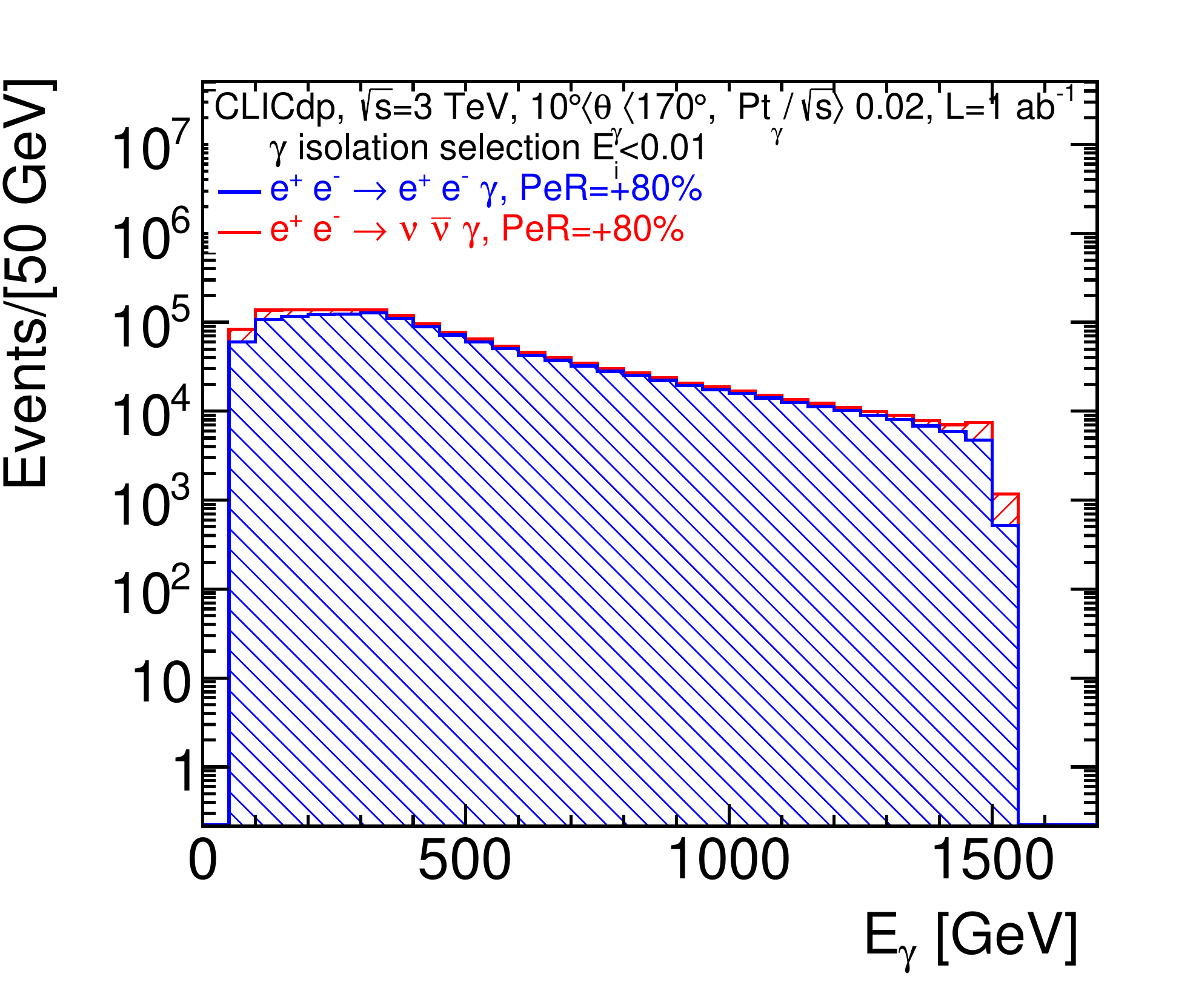}
    \caption{$dN/dE_{\gamma}$ spectrum with photon isolation selection $E_{i}<0.01$ ~~~~ }
    \label{fig:1DStackB2N3000}
  \end{subfigure}
  \hfill
  \begin{subfigure}[b]{0.48\textwidth}
    \includegraphics[width=\textwidth]{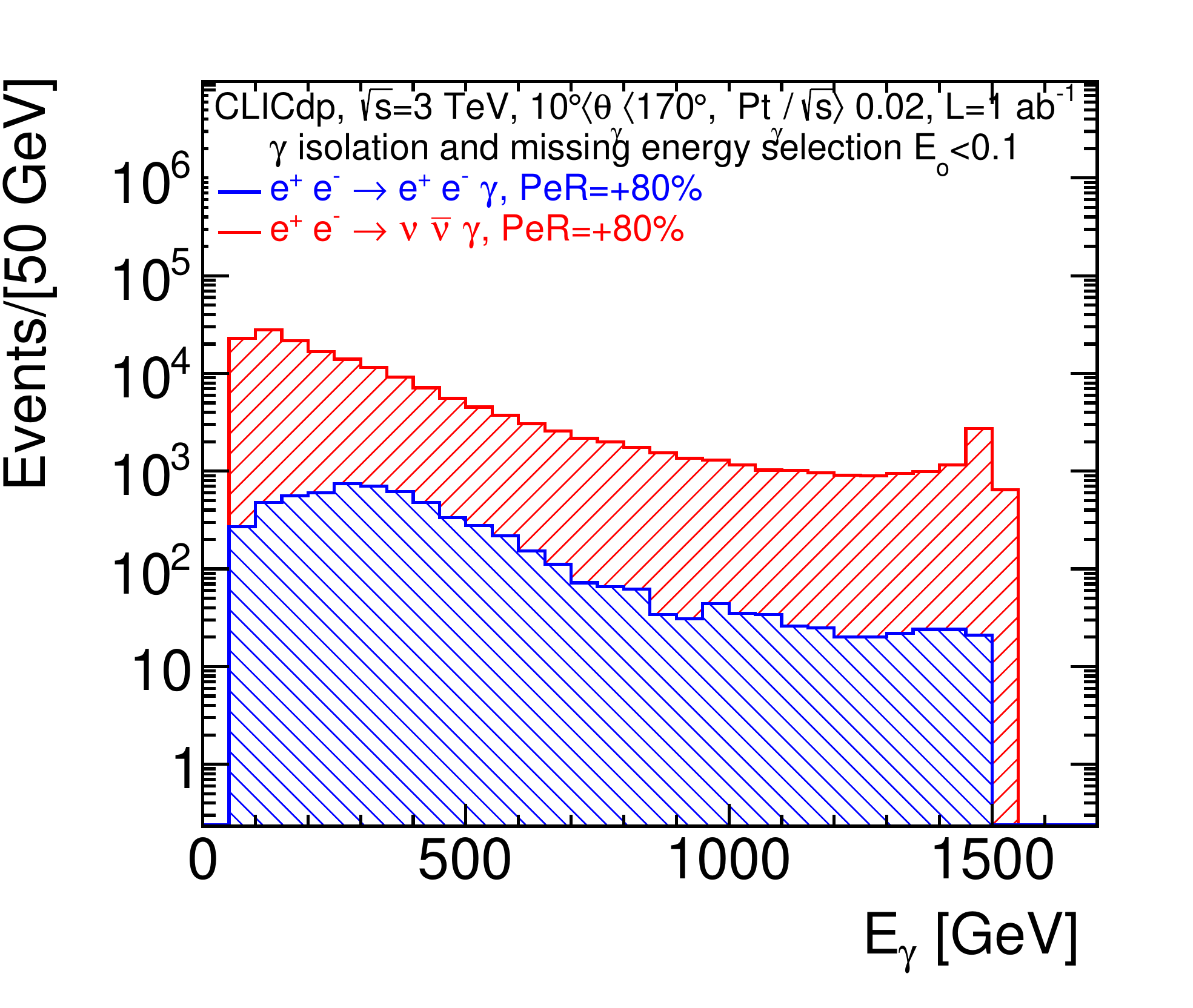}
    \caption{$dN/dE_{\gamma}$ spectrum with photon isolation selections $E_{i}<0.01$
     and missing energy selection $E_{o}<0.1$  }
    \label{fig:1DStackB23000}
  \end{subfigure}
  \caption{Right-handed 80\% polarised electron beam \mbox{ \Pem \Pep $\to$ \PGn \PAGn \PGg (\PGg)} and
   \mbox{ \Pem \Pep $\to$ \Pem \Pep \PGg (\PGg)} events with mono-photon selection
   \subref{fig:1DStackB2N3000} with photon isolation selection $E_{i}<0.01$ 
   \subref{fig:1DStackB23000}  with photon isolation selections $E_{i}<0.01$
     and missing energy selection $E_{o}<0.1$  }
\end{figure}

Figure~\ref{fig:1DStackB2N3000} 
shows for a right-handed 80\% polarised \Pem beam (PeR), the stacked histograms of the photon energy 
distribution of
\mbox{ \Pem \Pep $\to$ \PGn \PAGn (\PGg)}
and
\mbox{ \Pem \Pep $\to$ \Pem \Pep  (\PGg)}
events selected using
the photon isolation selection $E_{i}<0.01$.
%
Figure~\ref{fig:1DStackB23000} shows the photon energy distribution
for events selected using
the photon isolation selection $E_{i}<0.01$ 
and the missing energy selection $E_{o}<0.1$.

\subsection{95\% confidence level upper limit cross section calculation}
\begin{figure} [!ht]
  \centering
    \includegraphics[width=0.90\textwidth]{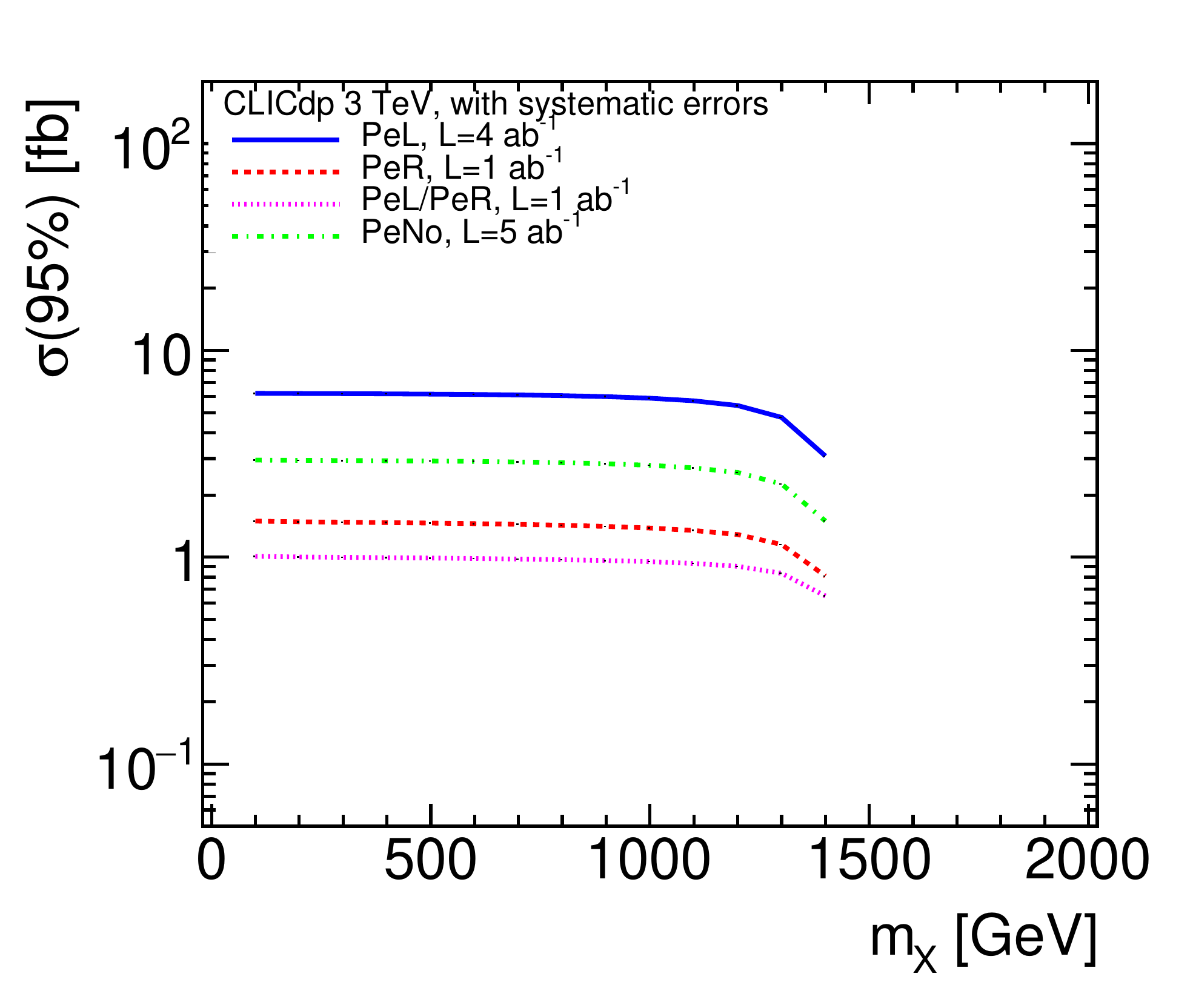}
  \caption{ 95\% confidence level upper limit cross section as a function of dark matter mass 
   for different polarisation and luminosity conditions.}
  \label{fig:XS95}
\end{figure}

The photon energy distributions, with photon isolation and missing energy selection, shown 
in Figure~\ref{fig:1DStackB13000} and ~\ref{fig:1DStackB23000}, together with 
the systematic uncertainties
listed in Table~\ref{tab:3000SystErrs}
are input to the 95\% confidence level upper limit cross section calculation
for the DM signal observation.
\begin{table} [!htbp]
\centering
\caption{Main systematic uncertainties at $\sqrt{s}$ = 3 TeV }
\begin{tabular}{ l c  }
\hline
 Systematic error source                            & value     \\ \hline
 $\nu  \bar{\nu} (\gamma) $ event selection                           & $2.0\times10^{-3}$     \\
 $ e^{+} e^{-} (\gamma) $ event selection                           & $5.0\times10^{-5}$     \\
 Luminosity measurement                           & $2.0\times10^{-3}$     \\
 Polarisation measurement                         & $2.0\times10^{-3}$     \\
\end{tabular}
\label{tab:3000SystErrs}
\end{table}
%
The 95\% confidence level upper limit cross sections are computed using the ratio of confidences
in the signal plus background to background hypothesis, so called ``CLS'' method~\cite{READ:A}.
For a counting experiment with a single channel 
\begin{eqnarray}
CL_{s}= {CL_{s+b} \over CL_{b}} = 
{{\sum_{n=0}^{b} {e^{-(b+s)} (b+s)^{n} \over {n!}} \over {\sum_{n=0}^{b} {e^{-(b)} (b)^{n} \over {n!}} } },
}  
\end{eqnarray}
%
where $b$ is the number of background events and
$s$ is the number of signal events.
%
For the process \mbox{\Pem \Pep $\to$ X X \PGg},
the angular distribution of the photon is independent of the process but
the photon energy spectrum depends on the dark matter mass. 
To compute the 95\% confidence level upper limit cross section as a function of the dark matter
mass $m_{X}$, for each  $m_{X}$ value, the number of background events b is computed
using 
%
\begin{eqnarray}
\label {equ:int}
b=\int_{E_{\gamma~min}}^{E_{\gamma~max}} {{\partial N} \over {\partial E}_{\gamma}} dE_{\gamma}
,~~\text{with}~E_{\gamma min}=\text{60~GeV~and}~E_{\gamma max}=\sqrt{s}/2 - m_{X}^{2}/\sqrt{s}.
\end{eqnarray}
$CL_{s}(PeR),~CL_{s}(PeL)$ and $CL_{s}(PeNo)$ are computed for right-handed, left-handed 
80\% polarised beams and without polarisation $(PeNo)$.  
The number of signal events excluded at 95\% CL, $s_{exc}$, is obtained for $CL_{s}$ >= 0.05.
%
%
To derive the limits using the right-handed and left-handed polarised photon energy distributions,
for each  $m_{X}$ value, the ratio $R_{b}$ and the error $\sigma(R_{b})$ are computed, 
\begin{eqnarray}
R_{b}= {b(PeL) \over b(PeR)},   
\end{eqnarray}
where $b(PeL)$ and $b(PeR)$ are the number of background events for left-handed and 
right-handed polarised beams, respectively.
The ratio $R_{b+s}$ is computed using
\begin{eqnarray}
R_{b+s}= {(b+s)(PeL) \over (b+s)(PeR)}.   
\end{eqnarray}
The expressions $(b+s)(PeL)$ and $(b+s)(PeR)$ are the number of background plus signal events for 
left-handed and right-handed polarised beams respectively and
assuming the polarisation dependance of dark matter Simplified Models.
The number of signal events excluded at 95\% CL, $s_{exc}$, is obtained for 
$R_{b}-R_{b+s} >= 2\times\sigma(R_{b})$.
The 95\% confidence level upper limit cross sections are
\mbox{$\sigma(95\%)$ = $s_{exc}/L$}, 
where 
$L$ is the integrated luminosity corresponding to the
polarisation condition.
Figure~\ref{fig:XS95} shows the 95\% confidence level upper limit cross section
as a function of the dark matter mass for different polarisation and luminisity conditions,
left-handed 80\% polarised \Pem beam, right-handed 80\% polarised \Pem beam,
using the ratios $R_{b}$ and $R_{b+s}$ for left-handed and right-handed 80\% polarised \Pem beams and 
without polarisation.
The lowest 
95\% confidence level upper limit cross section
is obtained using the ratios $R_{b}$ and $R_{b+s}$ 
for left and right-handed polarised \Pem beams.
The method using the ratios $R_{b}$ and $R_{b+s}$ is model dependant.
%
\subsection{Exclusion limits in simplified dark matter models}
\begin{figure} [!htbp]
  \centering
  \begin{subfigure}[b]{0.48\textwidth}
    \includegraphics[width=\textwidth]{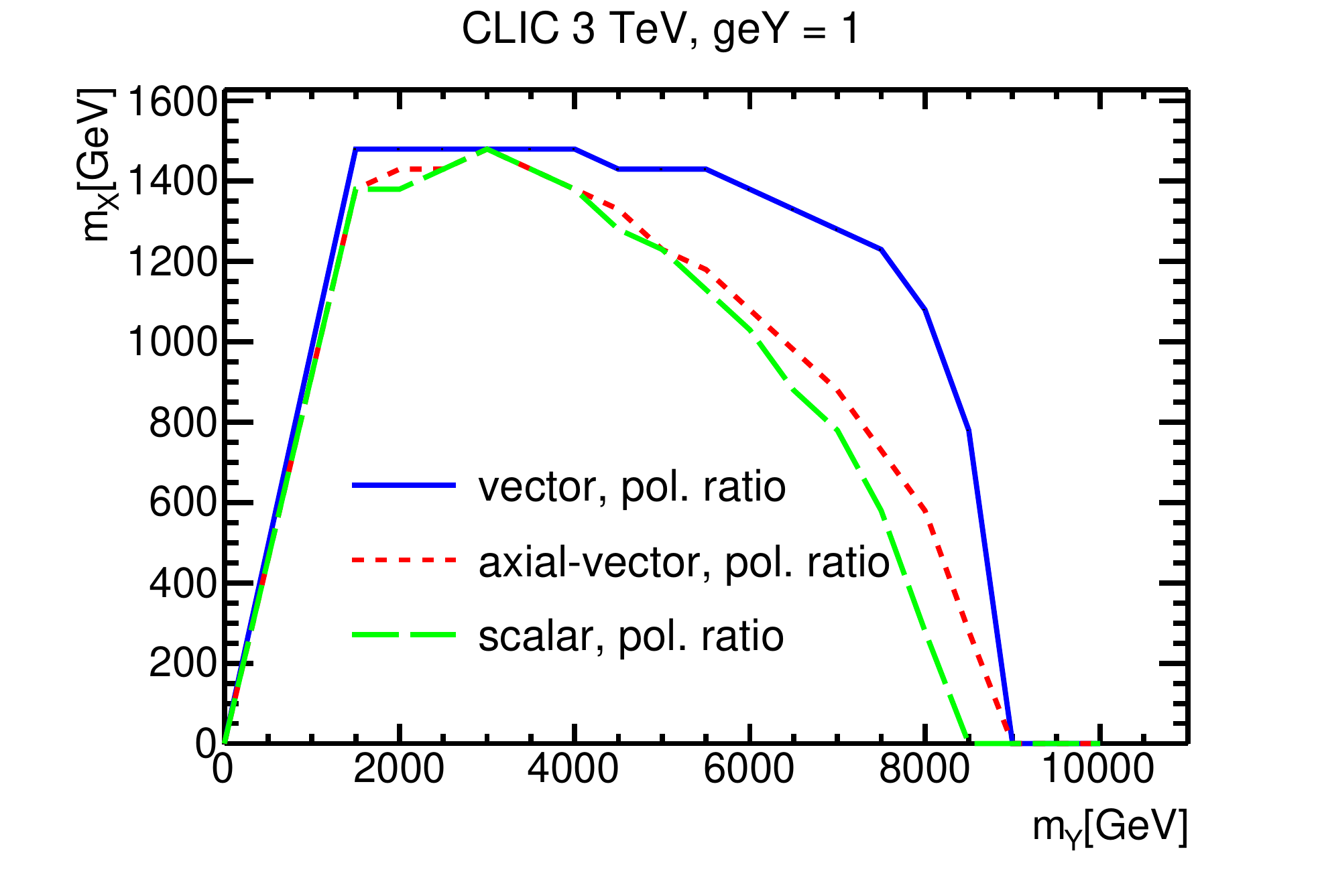}
    \caption{Exclusion limits in the mass plane $(m_{Y},m_{X})$ }
    \label{fig:Lim-geY1}
  \end{subfigure}
  \hfill
  \begin{subfigure}[b]{0.48\textwidth}
    \includegraphics[width=\textwidth]{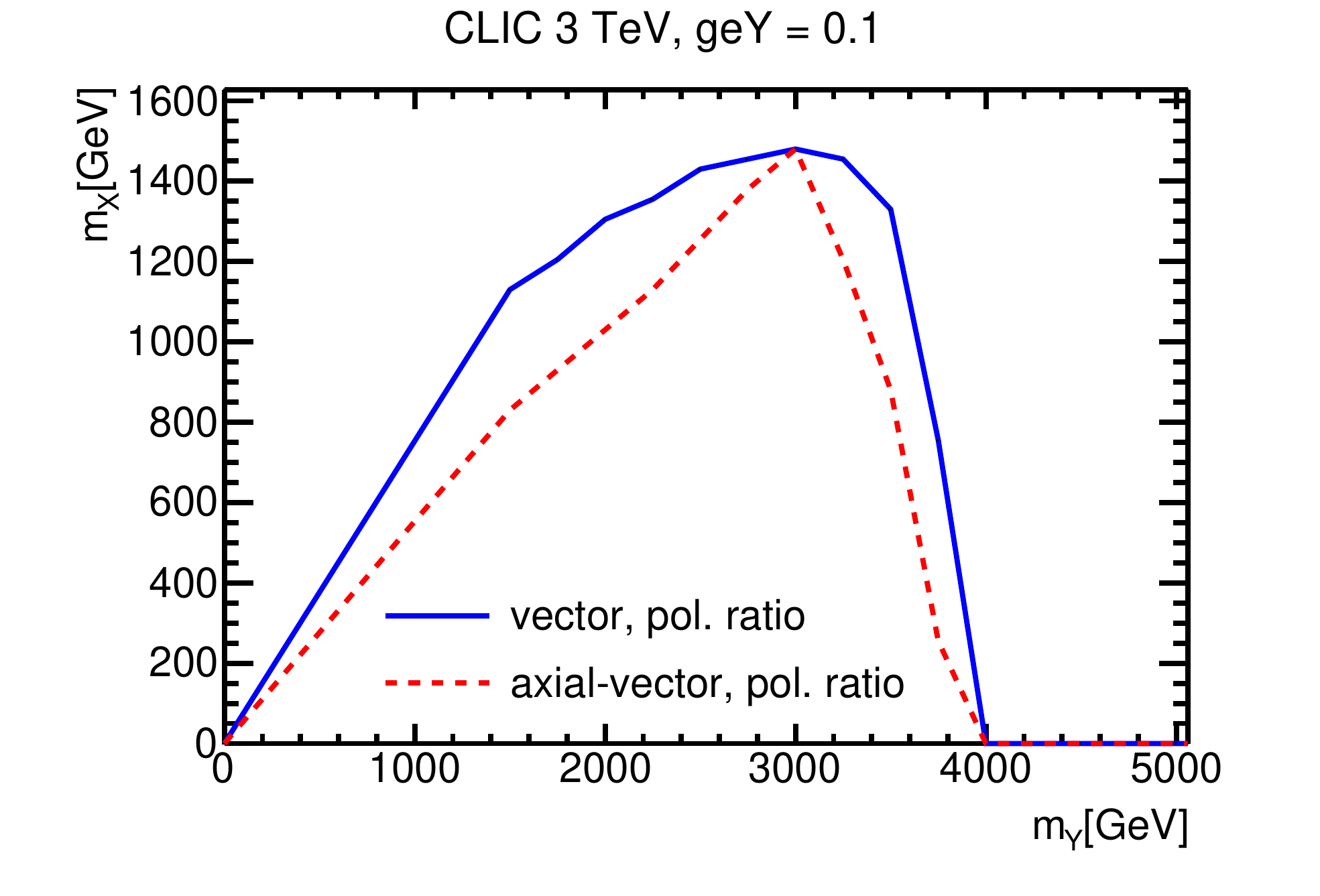}
    \caption{Exclusion limits in the mass plane $(m_{Y},m_{X})$ }
    \label{fig:Lim-geY0.1}
  \end{subfigure}
  \caption{ Exclusion limits in the mass plane $(m_{Y},m_{X})$ using the ratios $R_{b}$ and $R_{b+s}$
   and an integrated luminosity of 1 $\SI{}{ab^{-1}}$
   \subref{fig:Lim-geY1} $geY$=1  
   \subref{fig:Lim-geY0.1} $geY$=0.1  }
\end{figure}

In this study CLICdp focused on a subset of 
simplified dark matter models where the
mediator is exchanged in the s-channel. 
The parameters of these models are
the DM mediator Y type, vector (v) or axial-vector(a-v) or scalar(s),
the mediator mass $m_{Y}$, the e-e-mediator vertex coupling $geY$,
the DM mass $m_{X}$, and the mediator-DM-DM vertex coupling $gYX$.
For the cross section calculation the mediator width is fixed to 10 GeV.  
Limits are derived in the $(m_{Y},m_{X})$ plane using $\sigma(95\%)$ 
computed with $R_{b}-R_{b+s}$.
In the plane $(m_{Y},m_{X})$, in many points, the expected cross sections are computed.    
For each mediator mass $m_{Y}$, the limit in $m_{X}$ is the point   
where the cross section \mbox{$\sigma(m_{Y},m_{X})) >= \sigma(95\%)$}.
To compute the expected signal cross section, in each mass point,
requires generating events and applying the same event selection as for the
radiavite neutrino events. 
This is very time consuming, therefore a simplified method was considered.
In the simplified method the signal cross section is calculated using
the beam spectrum, the ISR function, one (ME) photon on which the generator cuts
are applied, but no ISR merging cut is applied.
For the same points the signal cross section is also computed using the 
same procedure as for the radiative neutrino events.
Applying the simplified method leads to cross sections which are are underestimated 
by 3\% to 4\% depending on the mass point. This leads to mass limits which 
are overestimated by 20 to 30 GeV. 
Figure~\ref{fig:Lim-geY1} shows, in the mass plane $(m_{Y},m_{X})$, the exclusion
limits computed for vector, axial-vector and scalar mediators with a coupling $geY$=1
and using the simplified cross section calculation.
For a light WIMP mass the exclusion range extends up to 9 TeV,
and WIMP masses close to half the centre-of-mass energy can be measured
for a large range of mediator masses.

Figure~\ref{fig:Lim-geY0.1} shows, in the mass plane $(m_{Y},m_{X})$, the exclusion
limits computed for vector and axial-vector mediators with a coupling $geY$=0.1.
For a light WIMP mass the exclusion range extends up to \mbox{4 TeV}, 
and WIMP masses of 1 TeV can be measured for mediator masses up to 3.5 TeV.
\section{Model discrimination and dark matter mass determination}
The model discrimination study and dark matter mass determination is done
using the simplified dark matter models with the following parameters: 
\begin {itemize}
\item DM mediator, v or a-v with $m_{Y}$=3.5 TeV and $\Gamma_{Y}$=10 GeV;
\item Coupling $geY$=1 or 0.5;
\item DM mass $m_{X}$ in the range between 200 GeV and 1.4 TeV;
\item Coupling $gYX$=1.
\end {itemize}
As can be seen on Figure~\ref{fig:Lim-geY0.1}, a mediator mass $m_{Y}$=3.5 TeV is close
to the exclusion limit, it leads to a challenging benchmark point. 
\subsection {Signal significance calculation}
\begin{figure} [!htbp]
  \centering
  \begin{subfigure}[b]{0.48\textwidth}
    \includegraphics[width=\textwidth]{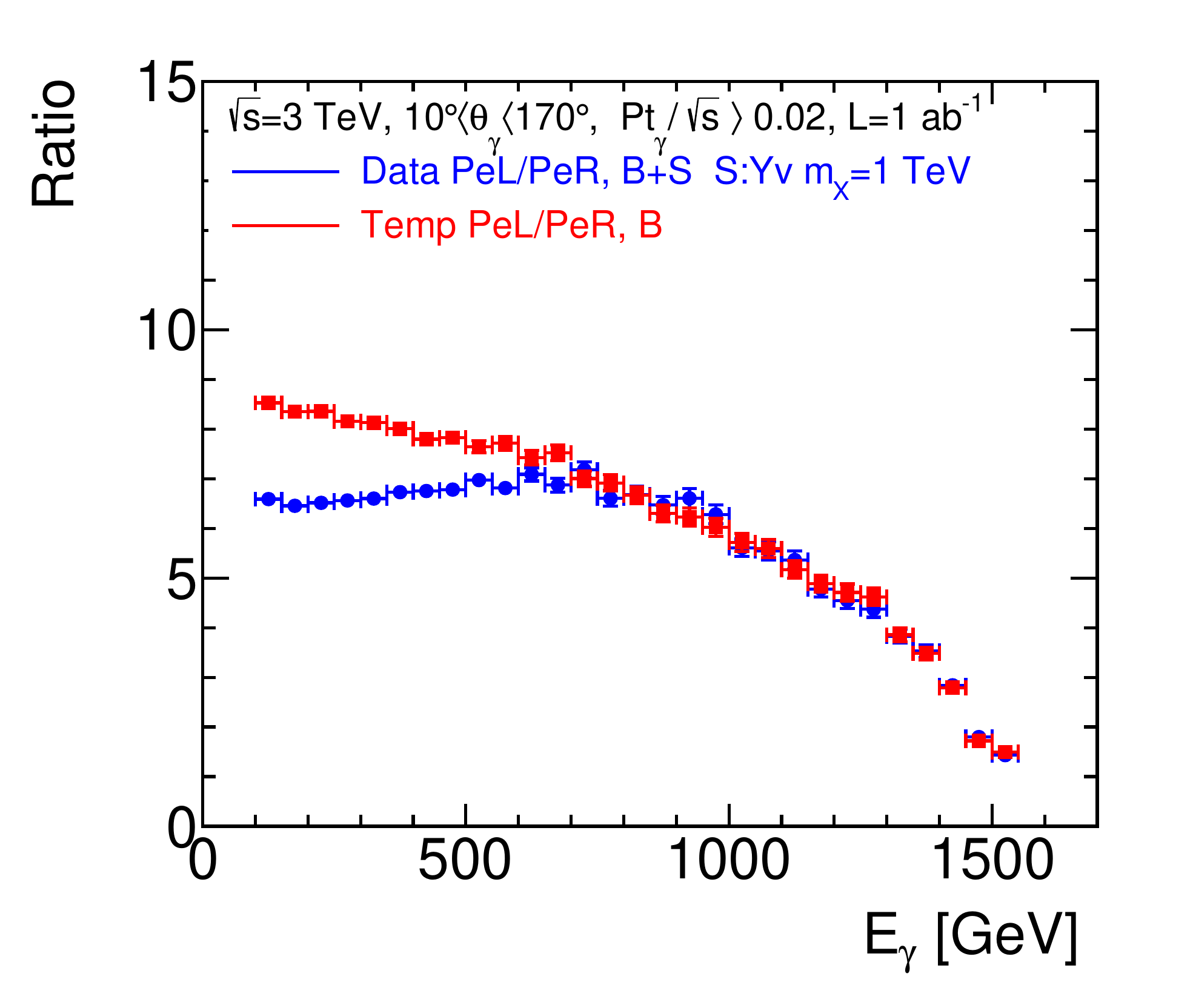}
    \caption{$dR_{B}/dE_{\gamma}$ background distribution and $dR_{D}/dE_{\gamma}$ pseudo-data 
     distribution for a
     vector mediator with \mbox{$m_{Y}$=3.5 TeV} and \mbox{$m_{X}$=1 TeV} }
    \label{fig:XSRatio}
  \end{subfigure}
  \hfill
  \begin{subfigure}[b]{0.48\textwidth}
    \includegraphics[width=\textwidth]{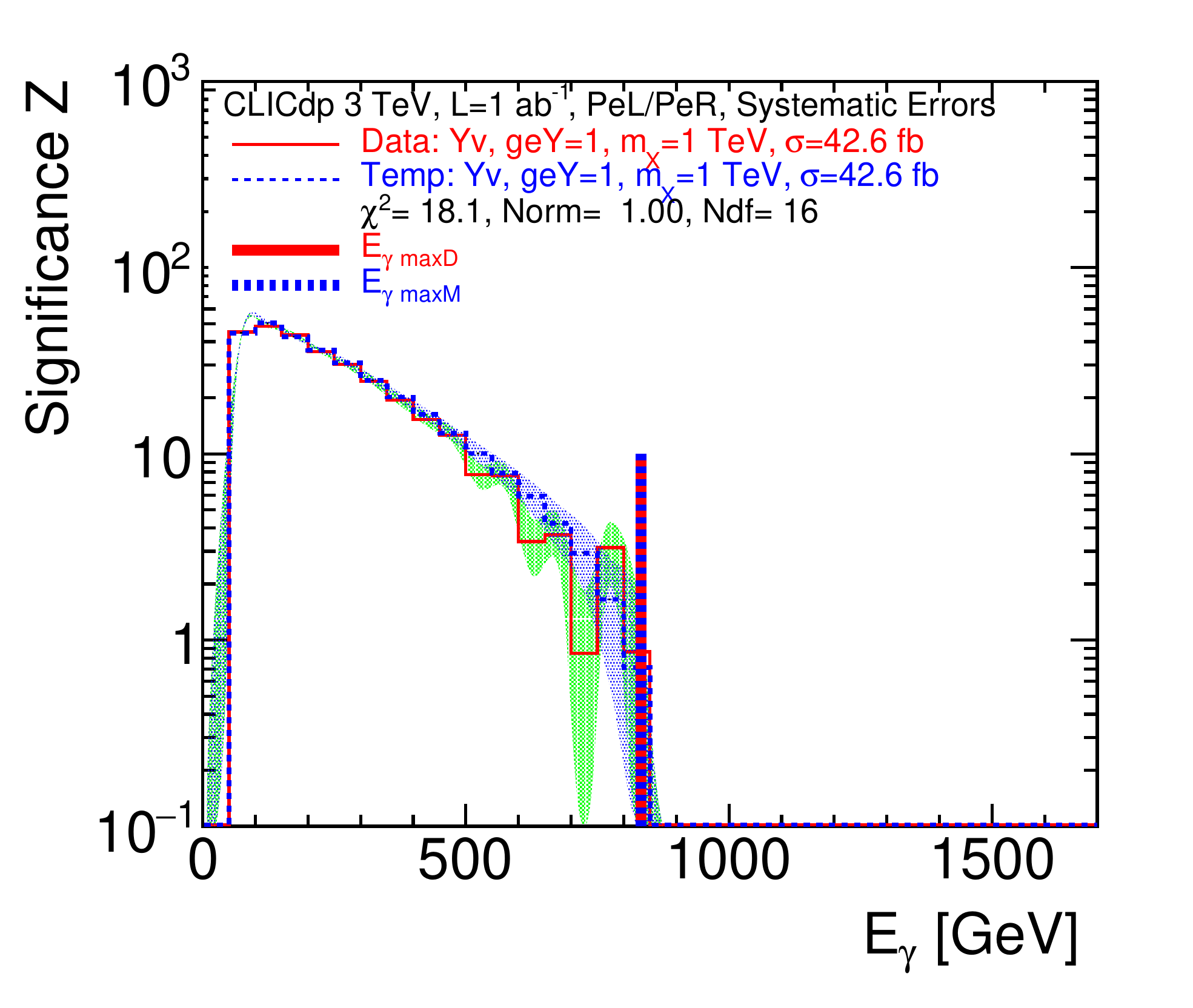}
    \caption{ $dZ/dE_{\gamma}$ distribusions: pseudo-data: vector mediator, 
    \mbox{$m_{Y}$=3.5 TeV} and \mbox{$m_{X}$=1 TeV}; template: vector mediator, 
     \mbox{$m_{Y}$=3.5 TeV} and \mbox{$m_{X}$=1} TeV }
    \label{fig:XZ}
  \end{subfigure}
  \caption{ 
   \subref{fig:XSRatio} Ratio of, left-handed over righ-handed, photon energy distribution
    for background 
    and for pseudo-data.
   \subref{fig:XZ} Pseudo-data and template significance $dZ/dE_{\gamma}$ distributions  }
\end{figure}

Figure~\ref{fig:XSRatio} shows, for background events, the ratio 
$dR_{B}/dE_{\gamma}$ of the photon energy distributions for left-handed
polarised \Pem beam over right-handed polarised \Pem beam.   
It shows also, for pseudo data events (background plus signal),
the ratio $dR_{D}/dE_{\gamma}$  of the photon energy distributions 
for left-handed polarised \Pem beam over right polarised \Pem beam. 
The signal events correspond to a vector mediator of mass $m_{Y}$=3.5 TeV with a coupling $geY$=1 
and a dark matter mass of 1 TeV.
%
\begin{eqnarray}
\label {equ:R}
{dR_{B} \over dE_{\gamma}}= {[dN_{B}/dE_{\gamma}]_{PeL} \over [dN_{B}/dE_{\gamma}]_{PeR}},~~~~
{dR_{D} \over dE_{\gamma}}= {[dN_{D}/dE_{\gamma}]_{PeL} \over [dN_{B}/dE_{\gamma}]_{PeR}},
\end{eqnarray}
where $dN_{B}/dE_{\gamma}$ is the photon energy distribution of background events,
and $dN_{D}/dE_{\gamma}$ is the photon energy distribution of pseudo-data events.
 
From these two distributions, the energy distribution of the
signal events $dN_{S}/dE_{\gamma}$ is
\begin{eqnarray}
\label {equ:S}
{dN_{S} \over dE_{\gamma}} = 
\biggl[{dN_{B} \over dE_{\gamma}}\biggr]_{PeR}
\biggl[{dR_{B} \over dE_{\gamma}} - {dR_{D} \over dE_{\gamma}}\biggr] 
\biggl[{dR_{D} \over dE_{\gamma}} -1\biggr].
\end{eqnarray}
For a discovery, the figure of merit is $1-CL_{b}$.
For a counting experiment with a single channel, 
$CL_{b}$ is computed in each $E_{\gamma}$ bin using 
\begin{eqnarray}
\label {equ:CLB}	
CL_{b}=  { \sum_{n=0}^{n_{obs}} {e^{-(b)} (b)^{n} \over {n!}} },  
\end{eqnarray}
where $b$ is the number of background events, $b=[dN_{B}/dE_{\gamma}]_{PeR}$,
and $n_{obs}$ is the number of signal plus background events, $s=dN_{S}/dE_{\gamma}$. 
The significance Z is derived 
\begin{eqnarray}
\label {equ:Z}	
Z= \sqrt{2}.Erf^{-1}[1-2(1-CL_{b})].
\end{eqnarray}
%
The uncertainty on the significance, $\delta Z$, is obtained using Toy Monte Carlo events 
to generate the distribution $dN/dZ$; a fit of $dN/dZ$ is performed to extract $\delta Z$. 
Using the same procedure, $dZ/dE_{\gamma}$ is computed for different
templates (vector, axial-vector) coupling to DM with different DM masses.
Figure~\ref{fig:XZ} shows the significance $Z$ as a function of $E_{\gamma}$ 
for a pseudo data sample and a template sample
corresponding to a vector mediator of mass $m_{Y}$=3.5 TeV
with a coupling $geY$=1 and a dark matter mass of 1 TeV;
the green band corresponds to $Z\pm 1\sigma$.
Figure~\ref{fig:XZ} shows also the significance $Z$ as a function of $E_{\gamma}$
for a template sample corresponding to a vector mediator of dark matter mass 1 TeV
and coupling $geY$=1.
For the template energy distributions the statistics is $10\times$ larger, to reduce fluctuations.
Before computing the significance distribution of the templates, the energy distributions 
are weighted to the expected luminosity.  
To compare the significance of the pseuso-data (ZData) and templates (ZTemp), 
a $\chi^{2}$ fit is performed with:
\begin{eqnarray}
\chi^{2} = \sum_{E_{\gamma=60 GeV}}^{E_{\gamma=1400 GeV}} { [(ZData(E_{\gamma})-ZTemp(E_{\gamma})]^{2} \over [\delta ZTemp(E_{\gamma})]^{2} }.
\end{eqnarray}
The $\chi^{2}$ fit is performed with the normalisation free (shape comparison) or with a normalisation using 
the pseudo-data and the template cross sections (absolute comparison).
%
%
%
\subsection {$\chi^{2}$ fit calculation check}
\begin{figure} [!htbp]
  \centering
  \begin{subfigure}[b]{0.48\textwidth}
    \includegraphics[width=\textwidth]{./plots/FNZEG_DMSPBP12_dmv1000dmv1000StatNoSmearLogMICDER_Bin_RB5_Ge1_F001.pdf}
    \caption{ $dZ/dE_{\gamma}$ distributions for pseudo-data and template samples with $geY$=1.}
    \label{fig:XZge1}
  \end{subfigure}
  \hfill
  \begin{subfigure}[b]{0.48\textwidth}
    \includegraphics[width=\textwidth]{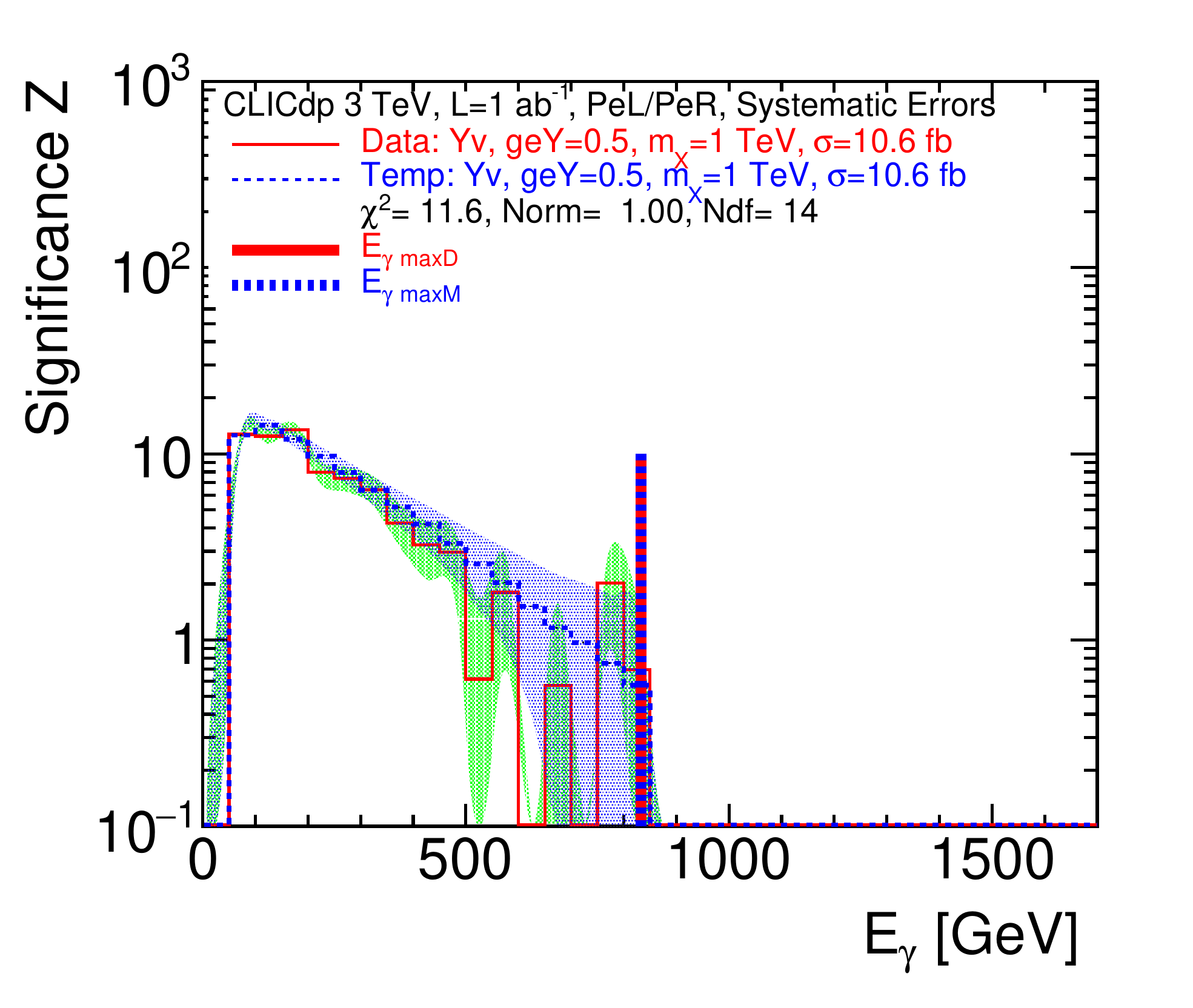}
    \caption{ $dZ/dE_{\gamma}$ distributions for pseudo-data and template samples with $geY$=0.5. } 
    \label{fig:XZge0.5}
  \end{subfigure}
  \caption{ Pseudo-data and template significance $dZ/dE_{\gamma}$ distributions for
           vector mediators with \mbox{$m_{Y}$=3.5 TeV} and \mbox{$m_{X}$=1 TeV}; 
   \subref{fig:XZge1} $geY$=1 
   \subref{fig:XZge0.5} $geY$=0.5  }
\end{figure}

Figure~\ref{fig:XZge1} shows the significance $Z$ as a function of $E_{\gamma}$ 
for a pseudo-data sample and a template sample corresponding to a vector mediator 
coupling to a dark matter of mass 1 TeV. 
For both samples the coupling $geY$=1.
The $\chi^{2}/Ndf$=1.13, $Ndf$ is the number of degrees of freedom.
Figure~\ref{fig:XZge0.5} shows the same distributions for pseudo-data and template samples
with a coupling $geY$=0.5. 
%
For a coupling $geY$=0.5 the cross sections are 4 times lower, the $\chi^{2}/Ndf$=0.82.
This check shows that the $\chi^{2}/Ndf$ is stable with respect to cross section changes. 
%
\subsection {$\chi^{2}$ fit and model discrimination}
\begin{figure} [!htbp]
  \centering
  \begin{subfigure}[b]{0.48\textwidth}
    \includegraphics[width=\textwidth]{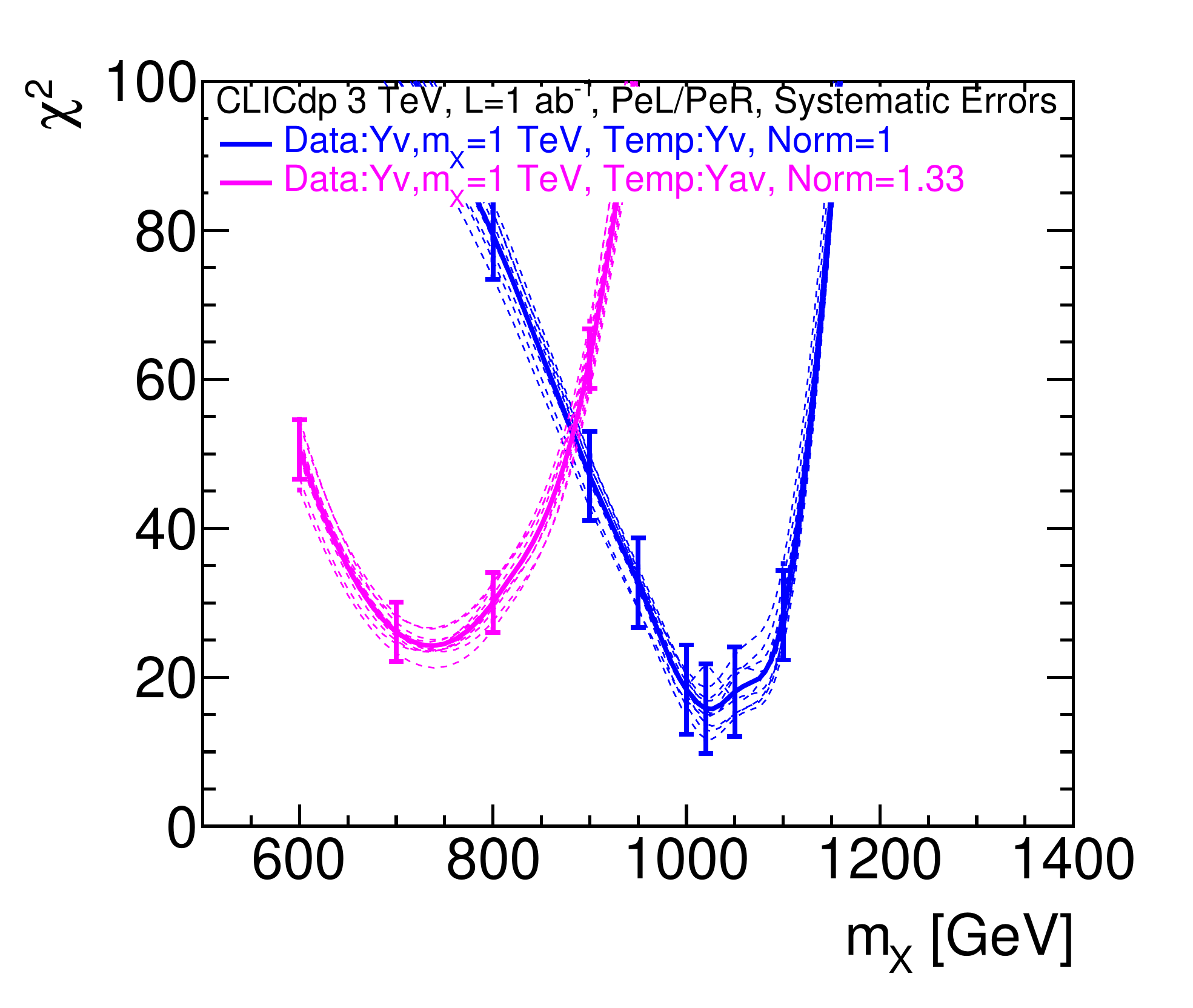}
    \caption{  $\chi^{2}$ fit values as a function of the dark matter mass, 
     pseudo-data: vector mediator with $m_{X}$=1 TeV, templates: vector and axial-vector mediator }
    \label{fig:Chi2v-v-and-v-a}
  \end{subfigure}
  \hfill
  \begin{subfigure}[b]{0.48\textwidth}
    \includegraphics[width=\textwidth]{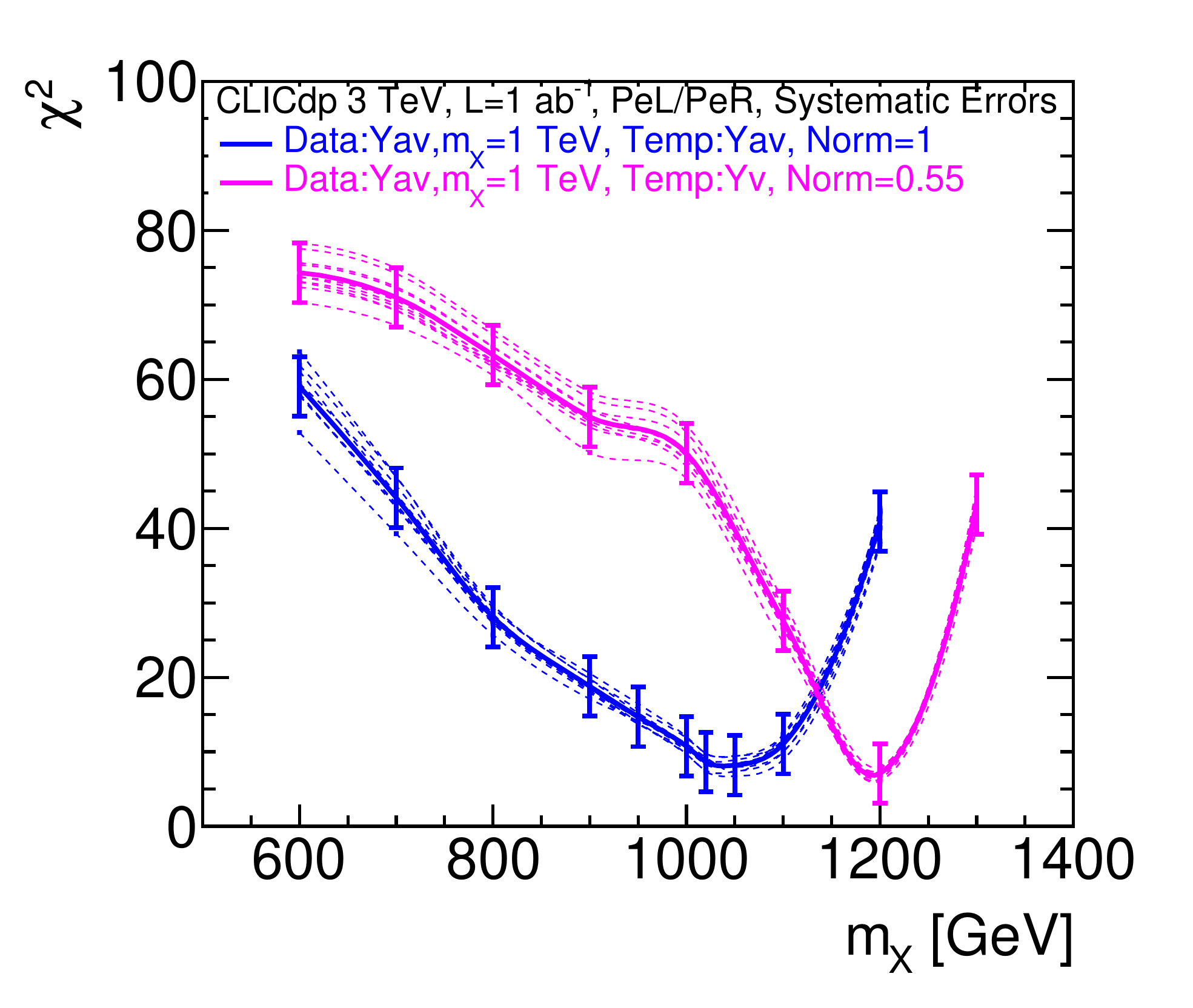}
    \caption{  $\chi^{2}$ fit values as a function of the dark matter mass, 
     pseudo-data: axial-vector mediator with $m_{X}$=1 TeV, templates: axial-vector and vector mediator }
    \label{fig:Chi2a-v-and-v-v}
  \end{subfigure}
  \caption{ $\chi^{2}$ fit values as a function of the dark matter mass, \mbox{$m_{Y}$=3.5 TeV},
   \subref{fig:Chi2v-v-and-v-a} Pseudo-data, vector mediator with $m_{X}$=1 TeV, 
    templates vector and axial-vector mediator
   \subref{fig:Chi2a-v-and-v-v} pseudo-data, axial-vector mediator with $m_{X}$=1 TeV, templates 
    axial-vector and vector mediator }
\end{figure}

Figure~\ref{fig:Chi2v-v-and-v-a} 
shows the $\chi^{2}$ fit values as a function of the dark matter mass.
The blue dotted lines correspond to $\chi^{2}$ values computed for
vector mediator pseudo-data samples
coupling to a dark matter mass of 1 TeV
and vector mediator templates coupling to dark matter masses ranging
between 200 GeV and 1.4 TeV.
Ten vector mediators pseudo-data samples are generated with the same conditions.
The blue full line corresponds to an average of the $\chi^{2}$ values of the ten samples.
The error bars drawn for these points correspond to $\sigma(\chi^{2})=\sqrt(2.Ndf)$.
%
%
The minimum $\chi^{2}$ value is 18 and the normalisation value is 1 for $m_{X}$ around 1 TeV.
The magenta dotted-lines correspond to $\chi^{2}$ values computed for vector mediator 
pseudo-data samples coupling to a dark matter mass of 1 TeV
and axial-vector mediator templates coupling to dark matter masses ranging
between 200 GeV and 1.4 TeV.
Ten axial-vector mediators pseudo-data samples are generated with the same conditions.
The magenta full line correspond to an average of the $\chi^{2}$ values of the ten samples.
The minimum $\chi^{2}$ value is 25 and the normalisation value is 1.33 for $m_{X}$ around 750 GeV.
For the pseudo-data sample and the template samples the coupling is $geY$=1.
The smallest $\chi^{2}$ values corresponds to the right mediator type but the $\chi^{2}$
separation is too small to 
discriminate the vector-vector hypothesis from the vector-axial-vector hypothesis.
Taking into account the normalisation allows discarding the vector-axial-vector hypothesis.
Different statistical tests are available to compare the shape of the two distributions. The 
Kolmogorov-Smirnov test was also performed on the same distributions, it leads to the same
conclusion, the shape comparison does not allow discriminating between different models. 
Figure~\ref{fig:Chi2a-v-and-v-v} 
shows the same distributions. 
The blue line and dotted lines correspond to $\chi^{2}$ values computed for
axial-vector mediator pseudo-data and template samples.
The minimum $\chi^{2}$ value is 8.6 and the normalisation value is 1 for $m_{X}$ around 1 TeV.
The magenta line and dotted-lines corresponds to a $\chi^{2}$ computed for an axial-vector mediator
pseudo-data sample and vector mediator templates.
The minimum $\chi^{2}$ value is 7.2 and the normalisation value is 0.55 for $m_{X}$ around 1.2 TeV.
The $\chi^{2}$ values are too close to allow the
discrimination between the axial-vector-vector hypothesis and the vector-vector hypothesis.
Taking into account the normalisation allows discarding the vector-vector hypothesis.
%
\subsection {$\chi^{2}$ fit and dark matter mass determination}
\begin{figure} [!htbp]
  \centering
    \includegraphics[width=\textwidth]{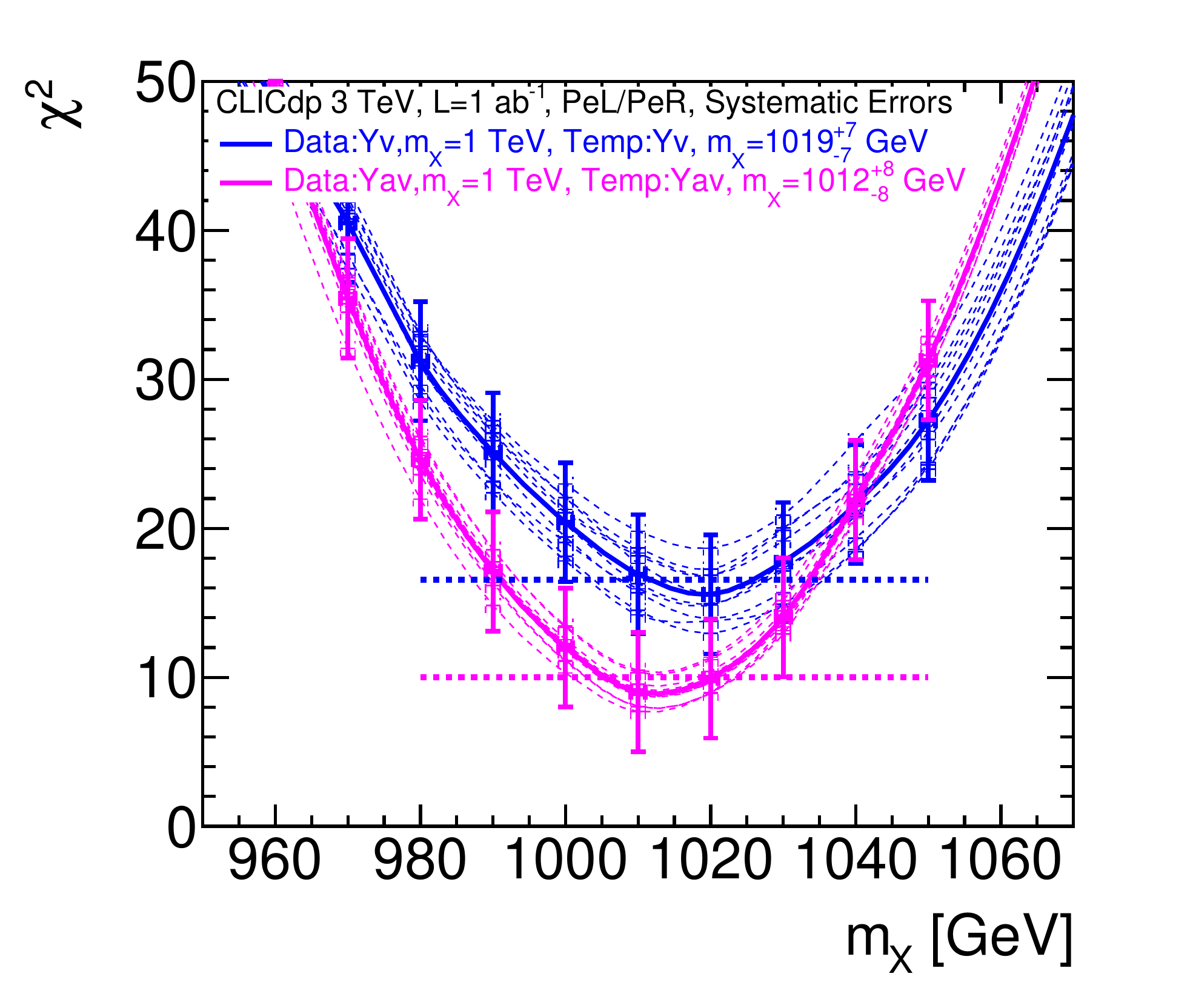}
  \caption{$\chi^{2}$ fit values as a function of the dark matter mass for pseudo-data 
   vector and axial-vector mediators with \mbox{$m_{Y}$=3.5 TeV} and $geY$=1 
   coupling to a dark matter mass of 1 TeV}
  \label{fig:Chi2v-v-and-a-a}
\end{figure}

Figure~\ref{fig:Chi2v-v-and-a-a} 
shows the $\chi^{2}$ fit values as a function of the dark matter mass.
The blue dotted line corresponds to a $\chi^{2}$ computed for 
vector mediator pseudo-data samples coupling to a dark matter mass of 1 TeV
and vector mediator templates coupling to dark matter masses ranging between 200 GeV and 1.4 TeV.
The ten vector mediators pseudo-data samples are all generated with the same conditions.
The blue full line correspond to an average of the $\chi^{2}$ values of the ten samples.
The error bars drawn for these points correspond to  $\sigma(\chi^{2})=\sqrt(2.Ndf)$.
The minimum $\chi^{2}$ value is 15.6 and the normalisation value is 1 for $m_{X}$ around 1 TeV.
For the pseudo-data samples and the template samples the coupling is $geY$=1.
The fit takes into account the shape of the $dZ/dE_{\gamma}$ distributions and the cross 
sections.
The minimum value of the $\chi^{2}$ gives the mass of the dark matter, 
the uncertainty on the mass, $\delta M$, corresponds to a $\chi^{2}$ variation of 1.
For the vector-vector hypothesis $m_{X}=1019^{+7}_{-7}$ GeV.

The magenta dotted lines corresponds $\chi^{2}$ values computed for a 
axial-vector mediator pseudo-data samples coupling to a dark matter mass of 1 TeV
and axial-vector mediator templates coupling to dark matter masses ranging
between 200 GeV and 1.4 TeV.
For the pseudo-data sample and the template samples the coupling is $geY$=1.
Ten axial-vector mediator pseudo-data samples are generated with the same conditions.
The magenta full line corresponds to an average of the $\chi^{2}$ values of the ten samples.
The minimum $\chi^{2}$ value is 9.0 and the normalisation value is 1 for $m_{X}$ around 1 TeV.
For the axial-vector-axial-vector hypothesis $m_{X}=1012^{+8}_{-8}$ GeV. 
%
%

\section{Summary}
To assess the CLIC physics potential for dark matter searches, 
the 95\% confidence level upper limit on the cross section is
computed as a function of the dark matter mass for different polarisation conditions.
For simplified dark matter models, the lowest 95\% confidence level upper limit cross section
is obtained using the ratios $R_{b}$ and $R_{b+s}$ for left-handed and right-handed 
polarised \Pem beams. 
Using the 95\% cross section as a function of the dark matter mass, exclusion limits are derived 
using dark matter Simplified Models.
For a coupling $geY$=1 and a light WIMP mass the exclusion range extends up to 9 TeV.
WIMP masses close to half the centre-of-mass energy can be measured
for a large range of mediator masses.
For a coupling $geY$=0.1 and a light WIMP mass the exclusion range extends up to 4 TeV, 
and WIMP masses of 1 TeV can be measured for mediator masses up to 3.5 TeV.
To discriminate between different mediator hypotheses coupling to different DM masses,  
$\chi^{2}$ fits of the differential distribution $dZ/E_{\gamma}$ of pseudo-data and templates
of different mediators coupling to DM with different DM masses are performed.  
The fit is using the shape of the differential distribution of the significance
$dZ/dE_{\gamma}$ and the cross sections.
It allows also the determination of the dark matter mass and of the uncertainty on the mass.
For a dark matter mass of 1 TeV the accuracy on the mass is 1\%,
without a statistically significant bias.
\section{Acknowledgments}
We are grateful to Daniel~Schulte and Dominik Arominski
for making available the \Pem \Pep, \PGg \PGg and \mbox{e \PGg} beam spectra at the various
$\sqrt{s}$ energies and to A. Wulzer for having designed and implemented the Simplified dark matter 
scalar model.
This work benefited from services provided by the ILC Virtual Organisation, supported by the
national resource providers of the EGI Federation.
This research was done using resources provided by the Open Science Grid,
which is supported by the National Science Foundation and the U.S.
Department of Energy's Office of Science.
%

\printbibliography[title=References]

\end{document}